\documentclass[revtex4]{emulateapj}
\usepackage{apjfonts}

\usepackage{lscape}
\usepackage{amsmath}
\usepackage{color}
\usepackage{ulem}

\def\ltsima{$\; \buildrel < \over \sim \;$}
\def\simlt{\lower.5ex\hbox{\ltsima}}
\def\gtsima{$\; \buildrel > \over \sim \;$}
\def\simgt{\lower.5ex\hbox{\gtsima}}

\newcommand{\oiii}{[O~III]$\lambda 5007$}
\newcommand{\ha}{H$\alpha$}
\newcommand{\hb}{H$\beta$}

\newcommand{\typeone}{X-ray type-1}
\newcommand{\typetwo}{X-ray type-2}
\newcommand{\typeonetwo}{X-ray type-1 + type-2}

\newcommand{\Lo}{$\log L_{\rm [O~III]}$}
\newcommand{\Lha}{$\log L_{\rm H\alpha}$}

\newcommand{\Locor}{$\log L_{\rm [O~III]}^{\rm cor}$}
\newcommand{\Lx}{$\log L_{\rm X}$}
\newcommand{\Lxh}{$\log L_{14-195}$}
\newcommand{\Lxs}{$\log L_{2-10}$}

\newcommand{\LoLxh}{$\log (L_{\rm [O~III]}/L_{14-195})$}
\newcommand{\LocorLxh}{log $(L_{\rm [O~III]}^{\rm cor}/L_{14-195})$}
\newcommand{\LoLxs}{$\log (L_{\rm [O~III]}/L_{2-10})$}
\newcommand{\LocorLxs}{log $(L_{\rm [O~III]}^{\rm cor}/L_{2-10})$}

\newcommand{\LhaLxs}{$\log (L_{\rm H\alpha}/L_{2-10})$}

\newcommand{\LoLha}{$\log (L_{\rm [O~III]}/L_{\rm H\alpha})$}

\newcommand{\lo}{$L_{\rm [O~III]}$}
\newcommand{\locor}{$L_{\rm [O~III]}^{\rm cor}$}
\newcommand{\lx}{$L_{\rm X}$}
\newcommand{\lxh}{$L_{14-195}$}
\newcommand{\lxs}{$L_{2-10}$}
\newcommand{\lha}{$L_{\rm H\alpha}$}
\newcommand{\lolx}{$L_{\rm [O~III]}/L_{\rm X}$}

\newcommand{\locorlx}{$L_{\rm [O~III]}^{\rm cor}/L_{\rm X}$}
\newcommand{\locorlxs}{$L_{\rm [O~III]}^{\rm cor}/L_{2-10}$}

\newcommand{\ergs}{erg cm$^{-2}$ s$^{-1}$}
\newcommand{\erg}{erg s$^{-1}$}

\newcommand{\nh}{$N_{\rm H}$}
\newcommand{\fscat}{$f_{\rm scat}$}
\newcommand{\ihost}{$i_{\rm host}$}

\lefthead{Ueda et al.}
\righthead{\oiii\ properties of hard X-ray selected AGNs}
\begin{document}

\title{
\oiii\ and X-ray Properties 
of a Complete Sample of Hard X-ray Selected AGNs in the Local Universe
}

\author{
Y. Ueda\altaffilmark{1},
Y. Hashimoto\altaffilmark{2},
K. Ichikawa\altaffilmark{1,3},
Y. Ishino\altaffilmark{1},
A.Y. Kniazev\altaffilmark{4,5,6},
P. V\"ais\"anen\altaffilmark{4,5}, 
C. Ricci\altaffilmark{1,7,8},
S. Berney\altaffilmark{9},
P. Gandhi\altaffilmark{10},
M. Koss\altaffilmark{9},
R. Mushotzky\altaffilmark{11},
Y. Terashima\altaffilmark{12},
B. Trakhtenbrot\altaffilmark{9},
M. Crenshaw\altaffilmark{13}
}
\altaffiltext{1}{Department of Astronomy, Kyoto University, Kyoto
606-8502, Japan}
\altaffiltext{2}{Department of Earth Sciences, National Taiwan Normal
University, No.\ 88, Sec.\ 4, Tingzhou Rd., Wenshan District, Taipei
11677, Taiwan R.O.C}
\altaffiltext{3}{National Astronomical Observatory, 2-21-1 Osawa, Mitaka, Tokyo 181-8588, Japan}
\altaffiltext{4}{South African Astronomical Observatory, PO Box 9, 7935 Observatory, Cape Town, South Africa}
\altaffiltext{5}{Southern African Large Telescope Foundation, PO Box 9, 7935 Observatory, Cape Town,  South Africa}
\altaffiltext{6}{Sternberg Astronomical Institute, Lomonosov Moscow
State University, Moscow, Russia}
\altaffiltext{7}{Pontificia Universidad Catolica de Chile, Instituto de Astrofisica,
Casilla 306, Santiago 22, Chile}
\altaffiltext{8}{EMBIGGEN Anillo, Concepcion, Chile}
\altaffiltext{9}{Department of Physics, Institute for Astronomy, ETH Zurich, Wolfgang-Pauli-Strasse 27, CH-8093 Zurich, Switzerland}
\altaffiltext{10}{School of Physics \& Astronomy, University of Southampton, Highfield, Southampton SO17 1BJ, UK}
\altaffiltext{11}{Department of Astronomy, University of Maryland, College Park, MD 20742-2421, USA}
\altaffiltext{12}{Department of Physics, Ehime University, Matsuyama 790-8577, Japan}
\altaffiltext{13}{Department of Physics and Astronomy, Georgia State University, 
25 Park Place, Suite 605, Atlanta, GA 30303, USA}

\begin{abstract}
We study the correlation between the \oiii\ and X-ray luminosities of
local Active Galactic Nuclei (AGNs), using a complete, hard X-ray ($>10$
 keV) selected sample 
in the Swift/BAT 9-month catalog.
From our optical spectroscopic observations
at the South African Astronomical Observatory and the literature, 
a catalog of \oiii\ line flux for all 103 AGNs at 
Galactic latitudes of $|b|>15^\circ$ is complied.
Significant correlations with
intrinsic X-ray luminosity (\lx ) are found both for observed (\lo) and 
extinction-corrected (\locor) luminosities, separately for X-ray unabsorbed
and absorbed AGNs. We obtain the regression form of
\lo\ $\propto L_{\rm 2-10\; keV}^{1.18\pm0.07}$ and
\locor\ $\propto L_{\rm 2-10\; keV}^{1.16\pm0.09}$ from the whole sample.
The absorbed AGNs with low ($<$0.5\%) scattering fractions in soft
X-rays show on average smaller \lolx\ and \locorlx\ ratios than the
other absorbed AGNs, while those in edge-on host galaxies do not.
These results suggest that a significant fraction of this population are
buried in tori with small opening angles.
By using these \lo\ vs.\ \lx\ correlations, the 
X-ray luminosity function of local AGNs (including Compton thick AGNs) 
in a standard population synthesis model gives
much better agreement with the \oiii\ luminosity function derived from
the Sloan Digital Sky Survey than previously reported. This confirms
that hard X-ray observations are a very powerful tool to find AGNs with
high completeness.

\end{abstract}


\section{Introduction}

In order to reveal the growth history of supermassive black holes
(SMBHs) in galactic centers, it is crucial to completely survey all
types of active galactic nuclei (AGNs) in the universe. According to the
unified scheme of AGNs \citep{ant93}, an SMBH is surrounded by a
bagel-shaped, dusty torus and only the viewing angle determines the
observable nature of an AGN; one sees type-1 and type-2 AGNs when the line
of sight is unblocked and blocked by the torus, respectively, which
causes dust extinction of optical lights from the accretion disk and
broad line region (BLR) and photoelectric absorption (plus Compton
scattering) of the primary X-ray emission. Basically, the unified scheme
seems fairly successful to explain many aspects of AGN phenomena. The
spectrum of the X-ray background indicates that a dominant population
of AGNs are type-2 (obscured) AGNs \citep[e.g.,][]{ued14}.
Hence, surveys only using the broad emission lines or soft X-rays could
easily miss the main population of AGNs.

Emission lines from the narrow-line region (NLR), which is located
outside the inner torus region, should be observable both from type-1
and type-2 AGNs unless the SMBH is entirely surrounded by the torus.
Thus, as long as the unified scheme holds, narrow emission lines induced by an
AGN, 
such as \oiii, have been 
considered to be a useful indicator of the AGN luminosity, even 
in Compton-thick AGNs whose ``observed'' X-ray flux below 10 keV is
significantly attenuated \citep[e.g.,][]{lamassa09}.
If, however, there is a wide scatter between the line luminosity and 
the intrinsic AGN luminosity, surveys based on the narrow lines
may be subject to strong selection effects. Also, optical and ultraviolet lines
are very sensitive to
extinction by interstellar dust in the host galaxy and by circumnuclear
dust that may be present around the NLR.
Note that
contamination from star-forming activities in the host galaxy may
become a problem to make a clean AGN sample based on the \oiii\ flux
\citep{sim05, tob14}.

Hard X-ray observations
at rest-frame energies above 10 keV 
are able to provide the least biased AGN samples against
obscuration thanks to their strong penetrating power, except for heavily
Compton-thick AGNs with column densities of log \nh\ $\simgt 25$
\citep{tue08}. From these surveys, AGNs with very low scattering
fractions in soft X-rays
have been discovered \citep{ued07}, many 
of which were missed in previous optical 
surveys because of their weak \oiii\ emission.
It has been suspected that the AGNs might be
buried in very geometrically-thick tori, although \citet{hon14} suggest that 
a part of them may be subject to interstellar absorption by the
host galaxy. In geometrically-thick tori with small opening angles, 
the AGN should have fainter
intrinsic \oiii\ luminosity relative to the hard X-ray luminosity
compared with classical Seyfert 2 galaxies, because much less of the
nuclear flux leaks out to ionize the NLR. An extreme case can be found
in ultra luminous infrared galaxies (ULIRGs) that contain buried AGNs
almost entirely surrounded by Compton-thick matter \citep{ima07,ich14}.

Thus, AGN selections using \oiii\ line and hard X-rays are considered
to be complementary to each other in detecting obscured
populations. It is therefore very important to study the correlations
between \oiii\ and hard X-ray luminosities so that we can compare the
statistical quantities of AGNs (such as luminosity function) obtained
from these different surveys, and evaluate the completeness and
cleanness of each selection.
For this study, we need to use a statistically complete sample of all
types of AGNs with well known properties.

Following early works by \citet{mul94} for Seyfert 1s and 2s and by
\citet{pol96} and \citet{alo97} for Seyfert 2s, several authors
have studied correlations between \oiii\ and hard X-ray luminosities, using
various samples of local AGNs
\citep[e.g.,][]{hec05,pan06,net06,mel08,lam09}.  \citet{hec05} and
\citet{mel08} use observed \oiii\ luminosities (hereafter \lo), while
the others use those corrected for extinction (hereafter \locor). Among
these works, only \citet{mel08}, who have more focus on the
[O IV] 25.89$\mu$m line, use an AGN sample based on hard X-ray
surveys above 10 keV, although the sample is not statistically
complete and is limited in number (40). The quantitative results of
the \oiii\ and X-ray luminosity correlation obtained so far have been
a little puzzling. From combined samples of type-1 and type-2 AGNs,
\citet{pan06} obtained a regression of the form $L_{\rm [O~III]}^{\rm
cor} \propto L_{\rm X}^{0.82\pm0.04}$,
whereas \citet{lam09} found an almost linear correlation of
$L_{\rm [O~III]}^{\rm cor} \propto L_{\rm X}^{0.98\pm0.06}$.

In this paper, we investigate the correlation between the \oiii\ and
X-ray luminosities, using a {\it complete} sample consisting of 103
objects at Galactic latitudes of $|b|>15^\circ$ in the Swift/BAT 9-month
hard X-ray survey \citep{tue08}. To follow-up sources in the southern
hemisphere, many of which did not have optical spectra, we conducted systematic
optical spectroscopic observations at the SAAO. Then we complement it
with a compilation from the literature, including \citet{win10a}, where
the optical spectra of Swift/BAT 9-month AGNs in the northern sky are
analyzed. Section~2 describes the sample, optical observations and data reduction,
and present the catalog of the \oiii\ flux together with those of narrow
\ha\ and \hb\ lines whenever available. In Section~3, we present the
results of correlation analysis between the \oiii\ (or narrow \ha ) luminosity
and intrinsic (de-absorbed) X-ray luminosity for different types of AGNs.
We then discuss the origin of
these correlations, and compare \oiii, \ha, and X-ray luminosity
functions of local AGNs in Section~4. The conclusions are summarized in
Section~5.  Throughout the paper, we adopt $H_{0} = 70$ km s$^{-1}$
Mpc$^{-1}$, $\Omega_{\rm M}=0.3$, and $\Omega_{\Lambda}=0.7$.

\section{The Optical Spectroscopy Data}

\subsection{Parent Sample}\label{sec:sample}

For our study, we utilize the Swift/BAT 9 month catalog \citep{tue08}
to define a complete sample of hard X-ray selected AGNs in the local
universe. The \citet{tue08} catalog contains 137 AGNs in total
excluding blazars at a flux limit of $2\times 10^{-11}$ erg cm$^{-2}$
s$^{-1}$ in the 14--195 keV band with detection significance above
4.8$\sigma$. To minimize the effects of extinction by Galactic
interstellar medium, we limit the sample to those located at high Galactic
latitudes of $|b|>15^\circ$ for our optical spectral studies.
We exclude Cen A, which is a very nearby object, and
SWIFT~J0350.1--5019, which likely is confused by two AGNs,
PGC 13946 and ESO 201-IG 004 \citep{ric15}. These selections leave 103
AGNs that constitutes our ``parent'' sample (hereafter ``Sample~A'').

The Swift/BAT AGNs are extensively followed-up by X-ray observatories
covering below 10 keV, such as Swift/XRT, XMM-Newton, Suzaku, and
Chandra. Key spectral parameters in our study are the absorption column
density (\nh) and the fraction of scattered component (\fscat) for absorbed
AGNs, 
which are often obtained by utilizing a partially covered absorber (or
its equivalent) model. 
Because there can be other soft X-ray components that are spatially
unresolved from the AGN emission,
the \fscat\ value determined in this way
is a upper limit to the true scattering fraction.
Here we basically adopt the results of spectral analysis
summarized in Table~1 of \citet{ich12}, which was largely based on
\citet{win09a} and was revised from (then) available Suzaku results for some targets. 
In our paper, we further revised their table by referring to 
later papers utilizing Suzaku data for more objects. Furthermore,
for sources whose spectral parameters were
not well constrained by using only the Swift/XRT data in \citet{win09a},
we update their spectral parameters according to 
\citet{ric15}, who perform uniform broad-band spectral
analysis in the 0.3--150 keV band by including Swift/BAT spectra for the
whole AGN sample of the Swift/BAT 70 month catalog. We also utilize the
70-months averaged, de-absorbed 2--10 keV flux of the primary continuum
listed in the \citet{ric15} catalog, as well as the 9-months averaged
14--195 keV flux in the original \citet{tue08} catalog.

We divide the sample into two types, X-ray unabsorbed AGNs (hereafter
``\typeone\ AGNs'') and absorbed AGNs (``\typetwo\ AGNs''), which have
absorptions of log \nh\ $<22$ cm$^{-2}$ and log \nh\ $\geq 22$,
respectively. Among \typetwo\ AGNs, we call those with \fscat\
$<0.5\%$ as low scattering-fraction AGNs (so-called ``new type'' AGNs),
a putative population of AGNs deeply buried by geometrically thick
tori. Due to our revision of the X-ray spectral parameters in the
original Swift/BAT 9 month catalog, the sample of low scattering-fraction 
AGNs has been also updated\footnote{No.\ 51 and 120 are
newly included in this sample while No.\ 4 and 86 are excluded.} from
that originally defined in \citet{ich12}. 

Table~1 list the targets of
Sample~A with their basic X-ray properties: source number in
\citet{tue08} ($^{**}$ are attached to the low scattering-fraction
AGNs), source name, redshift, \nh , \fscat, observed luminosity in the
14--195 keV band (9-month average), absorption-corrected 2--10 keV
luminosity (70-month average), and reference for the X-ray spectral
parameters.
Though not listed in Table~1, we also compile the information on the
inclination angle of the host galaxy, \ihost , using the HyperLeda database, which
are available for 98 AGNs
\footnote{Except for No.\ 29, 31, 116, 124, 136, and 151 in Table~1}. 
In addition, the black hole mass (and hence an
estimate of Eddington ratio) is available for 99 AGNs
\footnote{Except for No.\ 23, 53, 87, 120, and 149 in Table~1} 
from \citet{win09a}.

Figure~1 plots the host inclination against log \nh\ for Sample~A. In
all plots of our paper, the diagonal crosses correspond to
\typeone\ (X-ray unabsorbed) AGNs and the filled circles to \typetwo\ (X-ray
absorbed) AGNs, among which the open circles denote those with low
scattering fractions. 
As noticed, nine objects out of 10 with \ihost\ $> 85^\circ$
are \typetwo\ AGNs, rejecting 
the null hypothesis 
that X-ray absorption is independent of the host inclination 
at $>98\%$ confidence level.
This is expected as galactic interstellar matter could
produce an X-ray absorption of log \nh\ $> 22$ when viewed
edge-on, 
and is in agreement with the deficiency of nearly edge-on Seyfert 1
galaxies reported by \citet{kee80}.
Except for that, there is no correlation between \ihost\ and
\nh. These results are consistent with the random distribution of the
orientation angle of the torus (or the accretion disk) with respect to
that of the galactic plane, confirming previous findings \citep{sch01}. We note
that the \ihost\ distribution of the AGNs with low scattering fractions
in our sample are not concentrated at large values; more than half of
this population are free from absorption by interstellar matter along
the galactic disk, which therefore cannot account for their observed low
scattering fractions. 
Indeed, a KS test for the \ihost\ distribution between the low
scattering-fraction AGNs and the rest of 
\typetwo\ AGNs in Sample~A yields 
a matching probability of 0.53.
By considering the small sample size, this does not necessarily
contradict the statistical result by \citet{hon14}; they obtain more
edge-on dominant \ihost\ distribution of the same population based on
a slightly larger sample collected from the literature, although some
of their sample are revised in our paper (see Section~\ref{sec:sample}).
We can conclude that there are at least two origins for their low
scattering fractions, (1) intrinsic nature of the nucleus and (2)
interstellar absorption in the host galaxy.

\subsection{Optical Observations at SAAO and Data Reduction}

We performed optical spectroscopic observations of Swift/BAT AGNs
visible in the southern sky ($\delta$ $<-10^\circ$) by using the
SAAO 1.9-m telescope with the Cassegrain spectrograph during four
observation runs: 2007 July, 2008 January, 2008 August, and 2009
February, each consists of roughly 14 nights. In this paper, we focus
on sources 
in the Swift/BAT 9 month
catalog, although our observation targets at the SAAO also include
those in the Swift/BAT 22 month catalog, whose results will be reported
in \citet{kos15}. 
In total, the spectra of 38 AGNs have been analyzed in this work.

We used the 300 lines mm$^{-1}$ grating, blazed at 6000 \AA, covering
about 4400-7600 \AA, with a 2 arcsec slit-width placed on the center of
each galaxy, producing a spectral resolution of $\approx$5 \AA. The
integration is split into a series of 150 second exposures, added up to
a total integration time ranging from 750 to 3600 sec. Wavelength
calibration of the spectra was obtained from CuAr arc lamp exposures
taken during the same night. A flux calibration was obtained from
long-slit (with 6 arcsec slit-width) observations of spectrophotometric
standard stars. To derive the sensitivity curve, we fit the observed
spectral energy distribution of the standard stars with a low-order
polynomial.

The spectral line flux of \oiii, and those of narrow components of 
\ha\ and \hb\ were measured using IRAF task
{\it splot} 
from the co-added, dispersion corrected, and flux-calibrated spectra.
If the lines were not significantly detected, we then estimated their 
upper limits (3$\sigma$) from the fluctuation of the noise level. 
The line fluxes are corrected for reddening from the Milky Way, by using 
the $E(B-V)$ map by \citet{sch11} and the reddening curve by \citet{car89} with $R_V=3.1$.
Finally, we
approximately corrected these fluxes for the slit loss in the following way. 
For each dispersed spectrum, we projected the 4500--5500 \AA\ region
onto the spatial axis and measured its spread by fitting with a
gaussian. We then calculated the fraction contained within the slit by
assuming that the image is axisymmetric. By comparing the results of
the same target taken on different days when available, we estimate
that the flux uncertainties are typically of 0.1--0.2 dex, depending
on the quality of the spectrum. This is similar to general errors in
the \oiii\ fluxes reported by \citet{whi92} when they are measured with
small (2--4 arcsec) apertures.  In some cases of broad line AGNs, we
were unable to reliably measure the fluxes (nor the upper limits) of the narrow
components of \ha\ and \hb\ lines by 
separating them from the broad components.

\subsection{Catalog}\label{sec:catalog}

To complement the results from the SAAO observations, we gather \oiii,
\ha, and \hb\ fluxes in the literature for AGNs in the northern
sky. We mainly adopt the results summarized by \citet{win10a} except
for those with too large uncertainties, and refer to other references
\citep{mul94,bas99,xu99,lan07,kos15} for the rest. We obtain
constraints on the \oiii\ flux for all 103 AGNs (48 \typeone\ and 55
\typetwo\ AGNs) of the parent sample defined in Section~\ref{sec:sample}
(Sample~A), where one object (No. 49) does not show 
detectable \oiii\ emission and hence has only an upper limit.
Among them, 77 objects (31 \typeone\ and 46 \typetwo\ AGNs) 
have reliable flux measurements (not upper limits)
of both narrow \ha\ and \hb\ emission lines, 
$F_{\rm H\alpha}$ and $F_{\rm H\beta}$, or their flux ratios, constituting ``Sample B''.

Table~1 lists
the observed \oiii\ luminosity (\lo) along with 
the fluxes of \oiii, narrow \ha, and narrow \hb\ lines for 
Sample~A with the reference of the optical spectroscopic data.
For Sample~B, we also calculate
an extinction-corrected luminosity of \oiii\ (\locor ) 
from the Balmer decrement as
$$
L_{\rm [O~III]}^{\rm cor} = L_{\rm [O~III]} (\frac{F_{\rm H\alpha}/F_{\rm H\beta}}{3.0})^{2.94}, 
$$
following \citet{bas99}. When the $F_{\rm H\alpha}/F_{\rm H\beta}$ ratio is
smaller than 3.0,
we do not apply any correction. 
As discussed in \citet{hao05}, however, the intrinsic flux ratio
between \ha\ and \hb\ in the NLR of an AGN could be different from the value
assumed here, being subject to the gas density and radiative transfer
effects. Also, there is an uncertainty in the correction 
because the 
spatial distributions of the \oiii\ and Balmer line emitting regions 
may not be the same 
due to the clumpiness of the NLR (see Section~\ref{sec:origin}).
Thus, we
should regard these corrections only as approximation.

\section{Correlations between X-ray and Optical Line Luminosities}

\subsection{Regression Analysis between X-ray and \oiii\ Luminosities}
\label{sec:regression}

Figure~2 plots the correlation of the observed \oiii\ luminosity (\Lo)
against (a) the luminosity in the 14--195 keV band (\Lxh) or (b) that in
the 2--10 keV band (\Lxs), using Sample~A. Figure~3 shows the same but for
the \oiii\ luminosity corrected for extinction (\Locor), using
Sample~B. For each plot, we evaluate the strength of the
luminosity-luminosity and flux-flux correlations separately for \typeone,
\typetwo, and all (\typeonetwo ) AGNs; the resultant Spearman's rank
coefficients and Student's t-null significance levels are summarized in
Table~2. We also calculate the ordinary least square bisector regression
lines of the luminosity-luminosity correlation
with the form of $Y=a+bX$ where $Y$ is either \Lo\ or \Locor\ and $X$
is either \Lxh\ or \Lxs. 
The parameters and their 1$\sigma$ errors are listed in Table~2. 
The best-fit lines obtained from all AGNs are plotted in Figures~2 and
3. When Sample~A is used, we ignore
the two objects whose \oiii\ luminosities are upper limits, and restrict
the luminosity range above \Lxs $>41$.

As shown in Table~2, 
we find significant correlations at confidence levels of $>99\%$ between
all combinations of the \oiii\ and X-ray luminosities for any AGN
types. 
The flux-flux correlations are weaker but significant at $>90\%$
confidence levels; relatively weak correlation is obtained for
the \typeone\ AGN sample, most probably due to the narrow X-ray flux range
($F_{\rm X} \simeq 2\times10^{-11} - 3\times10^{-10}$ \ergs\ in the 14--195 keV band).

From the luminosity correlations for the entire AGN sample, we obtain $b \approx 1.2$ in
the regression line, 
which is significantly ($> 1\sigma$) different from 1.
This result is confirmed by the recent work based on a larger 
but less complete sample of Swift/BAT AGNs at $z>0.01$ by \citet{ber15}. 
We find that the slope for the \typeone\ AGNs is smaller than \typetwo\ AGNs, 
although consistent within errors, given the large scatter of the
correlations. 
The correlations with respect to \lxh\ and those to \lxs\ are found to
be similar except for the normalizations.
This is expected because absorption has a small effect on the
observed hard X-ray luminosity (\lxh) 
except for heavily Compton thick AGNs, 
and \lxs\ is corrected for absorption through the X-ray spectral
analysis. 

We compare our results on the \locor\ - \lxs\ correlation obtained from 
\typetwo\ AGNs with previous works. The slope we obtain, $b=1.26\pm0.13$
(i.e., \locor $\propto$ $L_{\rm 2-10}^{1.26\pm0.13}$), is
somewhat larger than 
that of \citet{lam09},
who derive $b=0.98\pm0.06$
from a sample consisting of X-ray and optically selected Seyfert 2
galaxies. 
To check the effects of sample incompleteness of Sample~B, we
perform regression analysis with the {\it asurv} software \citep{iso86}, 
by considering the lower limits of
\locor\ to be \lo\ for the objects excluded in Sample~B.
We find that the slope $b$ changes only by $\sim$0.01 compared with the
case obtained from Sample~B. Hence the sample incompleteness cannot explain the
difference of our result from \citet{lam09}.
The reason behind the discrepancy is not clear, but 
could be due to the different sample selections and
luminosity ranges. \citet{lam09} include a sample compiled by
\citet{pan06} from the Palomar optical spectroscopic survey, which
covers a lower luminosity range ($L_{\rm 2-10} < 10^{42}$ \erg )
than our sample.
In fact, \citet{pan06} obtain a much smaller slope,
$b=0.75\pm0.09$, from their Seyfert 2 sample including Compton thick
AGNs, whose intrinsic X-ray luminosities are simply estimated by
multiplying by a constant factor. The flatter slope than ours would be
explained if contamination of \oiii\ from star formation in the host
galaxy is more significant in lower luminosity AGNs (see
Section~\ref{sec:origin}). Another possibility is enhanced past
activity in the low luminosity AGNs, which are left with a higher
\oiii\ luminosity with respect to the current low X-ray activity.

\subsection{Averaged \oiii\ to X-ray Luminosity Ratio}

We calculate the 
error-weighted mean
value of the \oiii\ to X-ray luminosity
ratio and its standard deviation for different AGN types. 
Here we consider a systematic error of 0.2 dex in \Lo\ and 0.5 dex
in \Locor\ in addition to the errors listed in Table~1.
The results
are summarized in Table~3. Although we find that the best-fit
regression line is not linear ($b>1$), its effect can be checked by
calculating an averaged \Lx\ value in each sample, which is also
listed in Table~3. In fact, we confirm that it little affects the
following discussions.

As noticed from Table~3, we find that the mean ratio of observed
\oiii\ luminosity (\lo) to X-ray luminosity is significantly smaller
in \typetwo\ AGNs than in \typeone\ AGNs by $\approx$0.4 dex, using
Sample~A. This trend remains the same for the extinction-corrected
\oiii\ luminosity (\locor) obtained from Sample~B, although the
difference between \typeone\ and \typetwo\ AGNs is reduced to 
0.1--0.2
dex. 
The ``reduction'' is consistent with the fact that the mean
extinction-correction factor is larger in \typetwo\ AGNs 
($<$ \locor /\lo$>$ = $0.59\pm0.09$) than in \typeone\ AGNs 
($<$ \locor /\lo $>$ = $0.29\pm0.11$).
This indicates a higher degree of obscuration toward
the NLR in \typetwo\ AGNs, consistent with previous results 
\citep[e.g.,][]{dah88,mul94,mel08}. 
It may be explained 
if the torus, or its extended structure such as dusty outflow \citep{hon12}, is
large enough to block a part of the narrow-line region.

To investigate the nature of AGNs with low scattering fractions, 
we calculate the mean \oiii\ to X-ray luminosity ratios
for two subsamples of \typetwo\ AGNs, (1)
those with \fscat $<0.5\%$ and (2) those
hosted by edge-on galaxies (\ihost $>80^\circ$). The results are also listed
in Table~3. We find that the mean extinction-corrected \oiii\
to X-ray luminosity ratio of the low scattering-fraction AGNs is much smaller than that of
the total \typetwo\ AGN sample, while that of the edge-on galaxies does
not differ from it within uncertainties. 
A simple $\chi^2$ test shows that the difference of the mean value of \locorlxs\
between the low scattering-fraction AGNs
and the other \typetwo\ AGNs is significant at 
$>99.9$\% confidence level.
This is also noticeable from
Figure~4(a), where we plot the \locorlxs\ ratio against log \nh\ for Sample~B.

These results suggest that 
a significant fraction of low scattering-fraction AGNs are 
indeed buried
in a torus with very small opening angles as originally proposed by \citet{ued07}.
This population of AGNs could contribute to reduce the averaged \locorlx\ ratio in
the total \typetwo\ AGN sample compared with that of \typeone\
AGNs, because they are predominantly identified as \typetwo\ AGNs
due to the large covering fraction by the torus.
We can rule out the possibility that their low scattering fractions
are merely the result of deficiency of scattering gas in the
NLR. If this were the case, we should observe a similar fraction of 
low \lolx\ objects among the \typeone\ AGN sample. 
Figure~4(b) plots the \locorlxs\ ratio against ``X-ray Eddington
ratio'' (the 2--10 keV luminosity divided by the Eddington
luminosity) using objects with available black hole masses in
\citet{win09a}. No clear correlation is noticeable for the whole
sample. The low scattering-fraction AGNs do not always have high
Eddington ratios, while \citet{nog10} report a possible negative
correlation between the scattering fraction and Eddington ratio.  
Theoretically, deeply buried AGNs would be expected in the early growth phases of SMBHs with
relatively small masses (hence with low luminosities).
Thus, to further investigate the natures of this population, 
we need a larger sample of low luminosity AGNs.

\subsection{Correlations with Narrow \ha\ Line Luminosity}

In AGNs, intense narrow \ha\ and \hb\ lines are also produced from the
NLR. 
Hence, we also perform
regression analysis between the narrow \ha\ and X-ray luminosities, and
that between the narrow \ha\ and \oiii\ luminosities, in the same way as
done in Section~\ref{sec:regression}. For each analysis, we utilize
objects in Sample~A that have available flux measurements of \ha\ or
\oiii. The correlation plots are displayed in Figures~5 and 6,
respectively, together with the best-fit linear regression forms, which
are given in Table~2. We also calculate the mean and standard deviation
of the \LhaLxs\ ratio and the \LoLha\ ratio, which are summarized in
Table~3. 

We find that, for all AGNs, (1) \lha\ $\propto$ $L_{\rm
2-10}^{1.02\pm0.08}$ with a similarly large scatter ($\approx$0.6 dex)
to that seen in the \lo\ vs.\ \lxs\ correlation, and that (2) \lo\
$\propto$ $L_{\rm H\alpha}^{1.19\pm0.05}$ with a much smaller scatter
($\approx$0.3 dex). 
The slope of the \lha\ vs.\ \lxs\ correlation
obtained from the \typetwo\ AGNs, $1.04\pm0.10$, is
larger than that obtained by \cite{pan06} from their sample of 34
Seyfert 2s, $0.78\pm0.09$, which covers a lower luminosity range
($L_{\rm 2-10} < 10^{42}$ \erg ) than ours.
Even though here we use only the luminosity of the ``narrow'' component
of \ha , the regression slope and scatter between \lha\ and \lxs\ are
similar to those found between ``total'' \lha\ (i.e., that including the
broad component) and \lxs\ \citep[e.g.,][]{ho08}.

\section{Discussion}

\subsection{Origin of Correlation and Scatter between \oiii\ and Hard X-ray Luminosities}
\label{sec:origin}

Using so far the largest ($N > 100$), statistically complete sample
of hard X-ray ($E >$14 keV) selected AGNs in the local universe, we
determine the statistical properties between \oiii\ and hard X-ray
luminosities with the best accuracy. The linear regression form of \lo\
$\propto$ $L_{\rm 2-10}^{1.18\pm0.07}$ (\locor\ $\propto$ $L_{\rm
2-10}^{1.16\pm0.09}$) is obtained from the whole sample (see
Table~2). These results can be used as the reference for AGNs in the
luminosity range of \Lxs\ = $41-46$. We also find that the mean
luminosity ratio between \lo\ 
and \lxs\ of \typetwo\ AGNs is
significantly smaller than that of \typeone\ AGNs. The difference is
largely contributed by a population of low scattering-fraction
AGNs. Another important result is the very large variance in the \LoLxs\
ratio, corresponding to its standard deviation of
$\sim$0.5 in \typeone\ AGNs and 
$\sim$0.7 in \typetwo\ AGNs (see Table~3).

The non-linear correlation (i.e., $b\ne1$) between \lo\ and \lx\
may be explained by a combination of multiple effects. The first
effect is the luminosity dependence of the AGN spectral energy
distribution. The luminosity of the narrow lines is predominantly
determined by the continuum flux of ultraviolet photons responsible for
photo-ionization of the NLR gas, rather than the X-ray flux. Thus, if
the spectral slope $\alpha$ (for the flux density $F_\nu \propto
\nu^{-\alpha}$) between UV and hard X-rays above 2 keV is larger in
more luminous AGNs as suggested by \citet{sco14}, it works to make the
\oiii\ to X-ray luminosity correlation steeper. Secondly, according to
the luminosity-dependent unification model \citep{ued03,ric13}, the
opening angle of an torus increases with luminosity, thus making the
angular spread of the ``NLR cone'' larger in more luminous AGNs. This
also leads to increase $b$. The third effect is due to the luminosity
dependence of the NLR size in the radial direction, which is
proportional to $L^{0.33\pm0.04}$ \citep{sch03}. Thus, the actual size
of the NLR might be saturated in very luminous AGNs if the outer
radius exceeds the scale height of the host galaxy \citep{net04}. The fourth effect
is the contamination of \lo\ from star formation in low luminosity
AGNs, as mentioned in Section~\ref{sec:regression}. The last two effects make the
regression slope flatter than unity.

To better understand the origin of the observed luminosity correlations
and scatters between \lo\ (or \locor) and \lx, comparison with the \lha\
and \lo\ correlation is useful. The Balmer lines are emitted by
recombination as the result of photo-ionization, whereas the \oiii\ line
is emitted via collisional excitation in the heated gas. Thus, the
intensity ratio between \ha\ and \oiii\ depends on the physical
parameters, such as the ionization parameter and density
\citep{fer83}. Also, the \oiii\ line comes preferentially from gas with a
density of $\sim 10^6$ cm$^{-3}$ unlike the Balmer lines, which come
from a wide range of densities. In fact, detailed images of the NLR with
{\it Hubble} Space Telescope for a few low-$z$ objects
\citep[e.g.,][]{eva91,fis13} show that much of the \oiii\ flux comes from
clumpy structures. The effects of dust extinction inside the NLR,
which may not be correctly measured with the Balmer decrement,
makes it even more complex.
Hence, depending on how the NLR gas and dust is
distributed, non-linear correlation as well as a significant scatter in
the flux ratio between the Balmer lines and \oiii\ line would be also
expected.

The results for all AGNs, 
\lha\ $\propto$ $L_{\rm 2-10}^{1.02\pm0.08}$ 
and 
\lo\ $\propto$ $L_{\rm H\alpha}^{1.19\pm0.05}$,
show that the observed non-linear correlation
($b\approx1.2$) between \lo\ and \lx\ cannot be simply explained by a
single reason. In addition to the four possibilities listed above, it
is found that the non-linear correlation between \lo\ and \lha, 
which is determined by plasma physics, also plays a role. The fact that
the slope between \lha\ and \lxs\ is close to unity ($b=1.02\pm0.08$)
indicates that 
the third effect (luminosity dependence of the NLR
physical size) and/or the fourth effect (contamination by star formation)
must work to cancel the first and second effects. 
The fact that flatter slopes are found from the \typeone\ AGNs,
which are dominant in the largest luminosity range, suggest that the
third effect is more important. 

The large variation between \lo\ and \lx\ may be explained because the
optical emission lines from the NLR are a secondary indicator of the
intrinsic AGN luminosity in that
they do not directly come from near the SMBH 
and have strong dependence
on the geometry and size of the NLR, its
averaged density, clumpiness, and amount of dust.
In fact, a significant scatter of $\sim$0.4 dex between the \oiii\ luminosity and
the continuum luminosity at 5100 \AA is also reported in the SDSS
quasar sample \citep{she11}.
The presence of the low scattering-fraction AGNs accounts for
the larger scatter of the 
the \lolx\ ratio in \typetwo\ AGNs ($\sim$0.7 dex) than in 
\typeone\ AGNs ($\sim$0.5 dex), which could be understood 
in terms of variation in the geometry (cone angle) of the NLR.
Since the correlation between \lo\ and \lha\ is found to be 
tighter than that between \lo\ and \lx, the clumpiness of the NLR gas
and dust extinction effects
would not be the prime cause of the \lo - \lx\ scatter. 
Another
effect could be time variability; even though we utilize ``70-month''
averaged hard X-ray fluxes, the emission from the NLR reflects the
past AGN power averaged over $>10^2$ years. 

\subsection{Comparison of \oiii, \ha, and X-ray luminosity functions}

The luminosity function (LF) is one of the most important statistical
properties of AGNs. Utilizing an AGN sample selected from the {\it
Sloan} Digital Sky Survey (SDSS), \citet{hao05} determined \oiii\ and
\ha\ LFs of AGNs at $z \leq 0.15$ (they adopt emission-line luminosities
not corrected for extinction, and we follow the same procedure below.)
\citet{hec05} then compared the SDSS \oiii\ LF with an X-ray LF in the
3--20 keV band derived from the {\it RXTE} Slew Survey
by \citet{saz04}. They found that the X-ray LF significantly underpredicts
the \oiii\ LF when the mean luminosity ratio between \lx\ and \lo\
obtained from the RXTE AGN sample is assumed without considering the
scatter. On the basis of this result, they argue that X-ray surveys seem
to miss a significant fraction of AGNs, particularly Compton-thick AGNs.

Recently, \citet{ued14} determined the X-ray luminosity function of AGNs
including Compton thick AGN over a redshift range of $z=0-5$, using a
highly complete sample of X-ray selected AGNs. The local AGN sample from
the Swift/BAT survey is also utilized. Detection biases against
(mildly) Compton thick AGNs are taken into account to correctly estimate their
intrinsic number. Because heavily Compton thick AGNs with log \nh\ =
25--26 are difficult to detect even in the $E>10$ keV hard X-ray band, they
assume that the fraction of AGNs with log \nh\ = 25--26 is the same as
those with log \nh\ = 24--25. The X-ray luminosity function and
absorption distribution function are used as the basis of a standard
population synthesis model of the X-ray background \citep{ued14}. We
note that the X-ray AGN LF by \citet{saz04} may not be appropriate to adopt for
direct comparison with LFs in other wavelengths because (1) the original
X-ray LF by \citet{saz04} was unfortunately affected by an error in the count rate to
flux conversion (by a factor of 1.4; see \citealt{saz08} and
\citealt{ued11}), (2) even after correcting for that error, it
significantly underestimates other X-ray LFs of Compton thin AGNs \citep{ued11}, and
(3) Compton thick AGNs, which are difficult to detect in the 3--20
keV band, are not included.

Thus, it is very interesting to make comparison with the \oiii\ and
\ha\ LFs with the most up-to-date X-ray LF of local AGNs including
Compton thick AGNs, in order to understand the completeness and cleanness
of AGN selections in these different wavelengths. The red curve in Figures~7(a)
and 7(b) represent the best-fit \oiii\ and narrow \ha\ LFs in
\citet{hao05} (two power-law model, the sum of Seyfert 1s and 2s), after correcting both
luminosity and space density for the difference of the adopted Hubble
constant, from $H_{0} = 100$ km s$^{-1}$ Mpc$^{-1}$ \citep{hao05} to
$H_{0} = 70$ km s$^{-1}$ Mpc$^{-1}$ (our paper). The black curve
in each figure is a prediction for \oiii\ (or \ha) LF calculated from
the \citet{ued14} X-ray LF at $z=0$. Here we convert \lxs\ into \lo\ (or \ha)
with the best-fit linear regression form (Table~2) separately for
\typeone\ and \typetwo\ AGNs, and also consider the scatter around it by
assuming a gaussian distribution with the standard deviation listed in
Table~3. As the \citet{hao05} result is obtained from AGNs at $z \leq 0.15$, we
then multiply luminosity-dependent density evolution factors
\citep{ued14} at the mean redshift. The black dashed curves denote the
boundaries when both errors (1$\sigma$) in the mean and standard
deviation of \LoLxs\ (or \LhaLxs) are taken into account. For
comparison, we also plot the case when the standard deviation is set
to be zero (i.e., no scatter is considered) with the blue, dot-dashed
curve. 

As noticed from Figure~7(a), the \oiii\ (red) and X-ray (black) LFs are
roughly consistent with each other within a factor of $\sim$2 when we
take into account the uncertainties in the \lx\ to \lo\
conversion. Thus, the systematic ($\approx$4) underestimate of the
\oiii\ LF by the X-ray LF over a wide range of luminosity reported by
\citet{hec05} is now resolved. Rather, at \Lo\ 
$\simgt$40, 
the X-ray
LF outnumbers the \oiii\ LF, while statistical uncertainties in the
\oiii\ LF are large (a factor of $>$2) at \Lo\ 
$\simgt$41.6 
due to the limited sample size in the SDSS. Figure~7(b) shows even
better agreement between the \ha\ and X-ray LFs over a wider
luminosity range, although a similar discrepancy is noticed at \Lha\
$\simgt$41.
We note that it is important to consider the
scatter between the two luminosities when making the comparison of
LFs, as seen in the difference between the black solid curve (with
scatter) and blue dot-dashed curve (without scatter).

These results confirm that 
hard X-ray ($>10$ keV) observations are a
very powerful tool to find AGNs with high completeness, not missing a
dominant portion of the entire AGN population,
once biases against Compton-thick AGNs are properly corrected (see e.g., \citealt{mal09}).
For the correction, however, it is essential to obtain the
broad-band X-ray spectra covering up to, at least, a few tens of keV, 
with sufficiently good sensitivities.
If the discrepancy between the
\oiii\ (or \ha) and X-ray LFs at the high luminosity range is true, this
instead implies that the optical selection would miss some AGN populations.
The selection based on emission-line diagrams could be incomplete for
AGNs significantly contaminated by star formation; indeed,
\citet{win10a} show that a non negligible fraction of hard X-ray selected AGNs
could be optically classified as H~II galaxies, even though they are truly AGNs. 
Other candidates of
``optically missing'' AGNs are those deeply embedded in tori with almost
spherical geometry, in which no or little NLR is formed. They may be
similar to some of the low scattering-fraction AGNs in our sample whose
\oiii\ fluxes are very weak. If many of heavily Compton thick AGNs
assumed in the \citet{ued14} model correspond to this population, it would
partially account for the mismatch between the optical and X-ray LFs.

\section{Conclusions}

From our observations at the SAAO and the literature, we have compiled a
{\it complete} catalog of \oiii\ line flux for 103 hard X-ray selected
AGNs in the local universe located at $|b|>15^\circ$, together with
narrow \ha\ and \hb\ line fluxes (or their ratio) for a large fraction
($\sim$80\%) of the sample. The main conclusions are summarized below.

\begin{enumerate}

\item We detect significant correlations between \oiii\ (without or with
extinction correction) and X-ray luminosities independently from
\typeone\ AGNs (log \nh\ $<22$) and \typetwo\ AGNs (log \nh\ $\geq22$),
even though there is a large scatter in their luminosity ratio.  The
best regression forms obtained from the whole sample are: \lo\ $\propto
L_{\rm 2-10\; keV}^{1.18\pm0.07}$ and \locor\ $\propto L_{\rm 2-10\;
keV}^{1.16\pm0.09}$.

\item 
Absorbed AGNs with low scattering fractions in the X-ray spectra
show smaller \lolx\ and \locorlx\ ratios than the other absorbed ones.
This suggests that a significant part of low
scattering-fraction AGNs are buried in tori with small opening angles.

\item Significant correlations are also found between the \ha\ and X-ray
luminosities. The \oiii\ and \ha\ luminosities are more tightly
correlated than the \oiii\ - X-ray luminosity correlation.

\item 
The X-ray luminosity function of local AGNs in a standard population
synthesis model shows much better agreement with the \oiii\ luminosity
function derived from the SDSS than previously reported.
It rather predicts a larger number of AGNs than the \oiii\ selection
at \Lo\ $\simgt$40.
This confirms that hard X-ray ($>10$ keV) observations are a very
powerful tool to find AGNs with high completeness, 
once biases against
Compton-thick AGNs are properly corrected on the basis of the broad-band X-ray
spectra.

\end{enumerate}

\acknowledgments

This paper uses observations made at the South African Astronomical
Observatory (SAAO). Part of this work was financially supported by
Grants-in-Aid for Scientific Research 26400228 (Y.U.) and for JSPS Fellows
for Young Researchers (K.I.) from the Ministry of Education, Culture,
Sports, Science and Technology (MEXT) of Japan, and by the National
Science Council of Taiwan under the grants NSC 99-2112-M-003-001-MY2 and
NSC 102-2112-M-003-016 (Y.H.). P.V.\ and A.Y.K.\ acknowledge the support from the
National Research Foundation (NRF) of South Africa.

\clearpage
\LongTables
%\begin{landscape}
\tablefontsize{\tiny}
\begin{deluxetable}{lcccccccccccll}
\tabletypesize{\footnotesize}
\tablewidth{0pt}
\setlength{\tabcolsep}{0.01in}
\tablenum{1}
\tablecaption{
X-ray and Optical Emission-Line (\oiii, \ha, \hb) Properties 
of AGNs in the 9-month {\it Swift}/BAT Catalog
\label{tbl-1}}
\tablewidth{0pt}
\setlength{\tabcolsep}{0.03in}
\tablehead{
\colhead{No.} & \colhead{Object} & \colhead{$z$} &  
\colhead{$\log L_{14-195}$} & \colhead{$\log L_{2-10}$} & \colhead{$\log N_{\rm H}$} & \colhead{$f_{\rm scat}$} &
\colhead{$\log L_{\rm [O III]}^{\rm cor}$} &
\colhead{$\log L_{\rm [O III]}$} &
\colhead{$\log F_{\rm [O III]}$} & \colhead{$\log F_{{\rm H}\alpha}$} & \colhead{$\log F_{{\rm H}\beta}$} &
\multicolumn{2}{c}{References}\\
\colhead{} &\colhead{}&\colhead{}&
\colhead{(erg s$^{-1}$)}& \colhead{(erg s$^{-1}$)}& \colhead{(cm$^{-2}$)} & \colhead{}&
\colhead{(erg s$^{-1}$)}& 
\colhead{(erg s$^{-1}$)}& 
\colhead{(erg cm$^{-2}$s$^{-1}$)}&\colhead{(erg cm$^{-2}$
 s$^{-1}$)}&\colhead{(erg cm$^{-2}$ s$^{-1}$)}& \colhead{(X-ray)}&\colhead{(Optical)}\\
\colhead{(1)} &\colhead{(2)}&\colhead{(3)}&
\colhead{(4)}& \colhead{(5)} & \colhead{(6)}& \colhead{(7)}& 
\colhead{(8)}&
\colhead{(9)}&
\colhead{(10)}&\colhead{(11)}& \colhead{(12)}& \colhead{(13)} & \colhead{(14)}
}
\startdata
1$^{**}$& NGC 235A &	0.0222& 43.56& 43.22& 23.50& 0.003&  42.17$\pm$ 0.43&  41.23$\pm$ 0.10&  $-$12.82$\pm$ 0.10&  $-$13.17$\pm$ 0.10&  $-$13.96$\pm$ 0.10& (xa)& (oa)\\
2$^{**}$& Mrk 348 &	0.0150& 43.68& 43.30& 23.20& 0.004&  41.95$\pm$ 0.01&  41.07$\pm$ 0.01&  $-$12.64$\pm$ 0.01&  $-$13.02$\pm$ 0.01&  $-$13.80$\pm$ 0.01& (xb)& (ob)\\
3& Mrk 352 &	0.0149& 43.27& 42.74& 20.75& \nodata&  \nodata&  40.39$\pm$ 0.01&  $-$13.31$\pm$ 0.01&  \nodata&  \nodata& (xc)& (oc)\\
4& NGC 454 &	0.0121& 42.88& 42.17& 23.30& 0.030& $>$40.87&  40.41$\pm$ 0.20&  $-$13.11$\pm$ 0.20&  $-$13.13$\pm$ 0.20& $<$$-$13.77& (xa)& (oa)\\
5& Fairall 9 &	0.0470& 44.39& 44.15& 20.36& \nodata&  \nodata&  42.15$\pm$ 0.20&  $-$12.56$\pm$ 0.20&  \nodata&  \nodata& (xc)& (oa)\\
6& NGC 526A &	0.0191& 43.63& 43.07& 22.18& \nodata&  41.44$\pm$ 0.43&  41.32$\pm$ 0.10&  $-$12.59$\pm$ 0.10&  $-$13.08$\pm$ 0.10&  $-$13.59$\pm$ 0.10& (xd)& (oa)\\
7& NGC 612 &	0.0298& 43.81& 43.45& 24.05& 0.006&  40.09$\pm$ 0.92&  40.09$\pm$ 0.13&  $-$14.22$\pm$ 0.13&  $-$13.96$\pm$ 0.11&  $-$14.41$\pm$ 0.29& (xe)& (od)\\
8$^{**}$& ESO 297-G018 &	0.0252& 43.85& 43.67& 23.81& 0.003&  41.43$\pm$ 1.02&  40.84$\pm$ 0.04&  $-$13.33$\pm$ 0.04&  $-$13.44$\pm$ 0.21&  $-$14.12$\pm$ 0.27& (xf)& (od)\\
9& NGC 788 &	0.0136& 43.39& 43.22& 23.67& 0.007&  40.91$\pm$ 0.49&  40.91$\pm$ 0.18&  $-$12.71$\pm$ 0.18&  $-$13.29$\pm$ 0.01&  $-$13.03$\pm$ 0.16& (xd)& (oe)\\
10& Mrk 1018 &	0.0424& 44.17& 43.61&  0.00& \nodata&  41.71$\pm$ 0.65&  41.71$\pm$ 0.09&  $-$12.91$\pm$ 0.09&  $-$13.46$\pm$ 0.20&  $-$13.91$\pm$ 0.09& (xa)& (oe)\\
12& Mrk 590 &	0.0264& 43.77& 42.71& 20.43& \nodata&  42.07$\pm$ 0.12&  41.74$\pm$ 0.04&  $-$12.46$\pm$ 0.04&  $-$13.00$\pm$ 0.01&  $-$13.59$\pm$ 0.04& (xc)& (oe)\\
15& NGC 931 &	0.0167& 43.66& 43.41& 21.56& \nodata&  41.91$\pm$ 0.03&  41.14$\pm$ 0.01&  $-$12.66$\pm$ 0.01&  $-$13.07$\pm$ 0.01&  $-$13.81$\pm$ 0.01& (xd)& (od)\\
16& NGC 985 &	0.0430& 44.21& 43.78& 21.59& \nodata&  42.75$\pm$ 1.99&  41.92$\pm$ 0.01&  $-$12.72$\pm$ 0.01&  $-$13.00$\pm$ 0.01&  $-$13.75$\pm$ 0.68& (xc)& (od)\\
17& ESO 416-G002 &	0.0592& 44.42& 43.57& 20.43& \nodata&  \nodata&  41.63$\pm$ 0.20&  $-$13.29$\pm$ 0.20&  \nodata&  \nodata& (xc)& (oa)\\
18& ESO 198-024 &	0.0455& 44.27& 43.50& 21.00& \nodata&  \nodata&  41.22$\pm$ 0.20&  $-$13.46$\pm$ 0.20&  \nodata&  \nodata& (xc)& (oa)\\
20$^{**}$& NGC 1142 &	0.0289& 44.17& 43.88& 23.80& 0.003&  42.03$\pm$ 0.03&  41.07$\pm$ 0.01&  $-$13.21$\pm$ 0.01&  $-$13.30$\pm$ 0.01&  $-$14.10$\pm$ 0.01& (xf)& (od)\\
23& PKS 0326--288 &	0.1080& 44.84& 44.42& 23.82& 0.009&  43.58$\pm$ 0.85&  42.52$\pm$ 0.20&  $-$12.95$\pm$ 0.20&  $-$13.15$\pm$ 0.20&  $-$13.98$\pm$ 0.20& (xa)& (oa)\\
24& NGC 1365 &	0.0055& 42.67& 42.03& 24.65& 0.260&  40.98$\pm$ 0.01&  39.62$\pm$ 0.01&  $-$13.21$\pm$ 0.01&  (8.7)&  \nodata& (xg)& (of)\\
25& ESO 548-G081 &	0.0145& 43.19& 42.91& 20.00& \nodata&  41.94$\pm$ 0.08&  40.99$\pm$ 0.01&  $-$12.69$\pm$ 0.01&  $-$12.65$\pm$ 0.01&  $-$13.45$\pm$ 0.03& (xd)& (od)\\
28& 2MASX J03565655--4041453 &	0.0747& 44.51& 43.70& 22.66& 0.032&  42.56$\pm$ 0.43&  42.18$\pm$ 0.10&  $-$12.95$\pm$ 0.10&  $-$13.29$\pm$ 0.10&  $-$13.90$\pm$ 0.10& (xa)& (oa)\\
29$^{**}$& 3C 105 &	0.0890& 44.83& 44.32& 23.75& 0.003&  42.79$\pm$ 0.10&  41.59$\pm$ 0.01&  $-$13.70$\pm$ 0.01&  $-$14.10$\pm$ 0.01&  $-$14.99$\pm$ 0.03& (xa)& (oe)\\
31& 1H 0419--577 &	0.1040& 44.91& 44.60& 24.31& \nodata&  43.55$\pm$ 0.43&  43.16$\pm$ 0.10&  $-$12.28$\pm$ 0.10&  $-$12.64$\pm$ 0.10&  $-$13.26$\pm$ 0.10& (xh)& (oa)\\
32& 3C 120 &	0.0330& 44.45& 43.98& 21.20& \nodata&  \nodata&  41.86$\pm$ 0.01&  $-$12.54$\pm$ 0.01&  \nodata&  \nodata& (xd)& (og)\\
34& MCG --01-13-025 &	0.0159& 43.41& 42.54& 19.60& \nodata&  40.95$\pm$ 0.03&  40.73$\pm$ 0.01&  $-$13.02$\pm$ 0.01&  $-$13.21$\pm$ 0.01&  $-$13.76$\pm$ 0.01& (xc)& (oe)\\
36& XSS J05054--2348 &	0.0350& 44.24& 43.49& 23.47& 0.009&  41.65$\pm$ 0.43&  41.42$\pm$ 0.10&  $-$13.03$\pm$ 0.10&  $-$13.15$\pm$ 0.10&  $-$13.71$\pm$ 0.10& (xf)& (oa)\\
38& Ark 120 &	0.0323& 44.11& 43.79& 20.30& \nodata&  \nodata&  41.35$\pm$ 0.01&  $-$13.03$\pm$ 0.01&  \nodata&  \nodata& (xd)& (oc)\\
39& ESO 362-G018 &	0.0126& 43.26& 42.88& 23.43& 0.087&  \nodata&  40.88$\pm$ 0.20&  $-$12.67$\pm$ 0.20&  \nodata&  \nodata& (xd)& (oa)\\
40& PICTOR A &	0.0351& 43.80& 43.45& 20.78& \nodata&  \nodata&  41.57$\pm$ 0.20&  $-$12.89$\pm$ 0.20&  \nodata&  \nodata& (xd)& (oa)\\
45& NGC 2110 &	0.0078& 43.54& 43.17& 22.45& 0.048&  41.64$\pm$ 0.01&  40.36$\pm$ 0.01&  $-$12.77$\pm$ 0.01&  $-$12.66$\pm$ 0.01&  $-$13.57$\pm$ 0.01& (xd)& (ob)\\
47& EXO 055620--3820.2 &	0.0339& 44.14& 43.07& 22.41& 0.034&  \nodata&  41.46$\pm$ 0.10&  $-$12.97$\pm$ 0.10&  \nodata&  \nodata& (xd)& (oa)\\
49$^{**}$& ESO 005-G004 &	0.0062& 42.56& 41.93& 24.06& 0.003& \nodata& $<$38.63& $<$$-$14.30&  $-$13.80$\pm$ 0.01& $<$$-$15.39& (xf)& (oh)\\
50& Mrk 3 &	0.0135& 43.61& 43.35& 24.04& 0.009&  43.33$\pm$ 0.01&  42.31$\pm$ 0.01&  $-$11.31$\pm$ 0.01&  $-$11.65$\pm$ 0.01&  $-$12.47$\pm$ 0.01& (xi)& (oe)\\
51$^{**}$& ESO 121-IG028 &	0.0403& 44.03& 43.63& 23.31& 0.004& $>$40.86&  40.86$\pm$ 0.24&  $-$13.72$\pm$ 0.24&  $-$13.69$\pm$ 0.10& $<$$-$13.76& (xa)& (oa)\\
53& 2MASX J06403799--4321211 &	0.0610& 44.40& 43.51& 23.00& $<$0.011& $>$41.74&  41.74$\pm$ 0.20&  $-$13.21$\pm$ 0.20&  $-$13.19$\pm$ 0.20& $<$$-$13.33& (xa)& (oa)\\
55& Mrk 6 &	0.0188& 43.72& 43.09& 20.76& \nodata&  42.38$\pm$ 1.56&  42.38$\pm$ 0.01&  $-$11.52$\pm$ 0.01&  $-$11.97$\pm$ 0.01&  $-$12.41$\pm$ 0.53& (xa)& (oe)\\
56& Mrk 79 &	0.0222& 43.72& 43.19& 19.78& \nodata&  41.96$\pm$ 0.12&  41.96$\pm$ 0.02&  $-$12.09$\pm$ 0.02&  $-$12.66$\pm$ 0.01&  $-$13.11$\pm$ 0.04& (xc)& (oe)\\
60& Mrk 18 &	0.0111& 42.93& 41.82& 23.26& 0.030&  40.56$\pm$ 0.28&  40.56$\pm$ 0.11&  $-$12.88$\pm$ 0.11&  $-$12.72$\pm$ 0.01&  $-$12.98$\pm$ 0.09& (xd)& (oe)\\
61& 2MASX J09043699+5536025 &	0.0370& 44.03& 43.31& 20.78& \nodata&  41.91$\pm$ 0.09&  41.63$\pm$ 0.03&  $-$12.87$\pm$ 0.03&  $-$12.95$\pm$ 0.01&  $-$13.52$\pm$ 0.03& (xd)& (oe)\\
62& 2MASX J09112999+4528060 &	0.0268& 43.69& 43.16& 23.52& 0.006&  40.98$\pm$ 0.11&  39.68$\pm$ 0.01&  $-$14.54$\pm$ 0.01&  $-$14.49$\pm$ 0.01&  $-$15.41$\pm$ 0.04& (xd)& (oe)\\
64& 2MASX J09180027+0425066 &	0.1560& 45.31& \nodata& 23.05& 0.013&  42.59$\pm$ 0.01&  42.22$\pm$ 0.01&  $-$13.60$\pm$ 0.01&  $-$14.08$\pm$ 0.01&  $-$14.68$\pm$ 0.01& (xd)& (oe)\\
65& MCG --01-24-012 &	0.0196& 43.60& 43.24& 22.81& 0.005&  41.10$\pm$ 0.15&  41.10$\pm$ 0.08&  $-$12.83$\pm$ 0.08&  $-$13.14$\pm$ 0.01&  $-$13.46$\pm$ 0.05& (xa)& (oe)\\
66& MCG +04-22-042 &	0.0323& 43.99& 43.46& 20.59& \nodata&  42.12$\pm$ 0.62&  42.12$\pm$ 0.20&  $-$12.26$\pm$ 0.20&  $-$12.57$\pm$ 0.01&  $-$12.72$\pm$ 0.20& (xc)& (oe)\\
67& Mrk 110 &	0.0353& 44.19& 43.86& 20.20& \nodata&  42.71$\pm$ 0.59&  42.29$\pm$ 0.19&  $-$12.17$\pm$ 0.19&  $-$12.47$\pm$ 0.01&  $-$13.09$\pm$ 0.19& (xd)& (oe)\\
68& NGC 2992 &	0.0077& 42.94& 41.93& 22.08& 0.524&  42.51$\pm$ 0.43&  40.76$\pm$ 0.10&  $-$12.36$\pm$ 0.10&  $-$12.48$\pm$ 0.10&  $-$13.55$\pm$ 0.10& (xd)& (oa)\\
69& MCG --05-23-016 &	0.0085& 43.55& 43.21& 22.20& \nodata& $>$41.18&  40.64$\pm$ 0.10&  $-$12.56$\pm$ 0.10&  $-$12.85$\pm$ 0.10& $<$$-$13.50& (xd)& (oa)\\
70& NGC 3081 &	0.0080& 43.09& 42.96& 23.99& 0.006&  41.66$\pm$ 0.43&  41.32$\pm$ 0.10&  $-$11.83$\pm$ 0.10&  $-$12.38$\pm$ 0.10&  $-$12.97$\pm$ 0.10& (xe)& (oa)\\
71& NGC 3227 &	0.0039& 42.63& 42.05& 22.24& 0.148&  41.26$\pm$ 0.01&  40.44$\pm$ 0.01&  $-$12.09$\pm$ 0.01&  $-$12.54$\pm$ 0.01&  $-$13.29$\pm$ 0.01& (xd)& (ob)\\
72& NGC 3281 &	0.0107& 43.27& 42.69& 23.94& 0.019&  41.42$\pm$ 0.85&  41.05$\pm$ 0.20&  $-$12.36$\pm$ 0.20&  $-$12.71$\pm$ 0.20&  $-$13.31$\pm$ 0.20& (xd)& (oa)\\
75$^{**}$& Mrk 417 &	0.0328& 43.95& 43.73& 23.93& 0.002&  41.19$\pm$ 0.05&  40.92$\pm$ 0.01&  $-$13.48$\pm$ 0.01&  $-$13.76$\pm$ 0.01&  $-$14.33$\pm$ 0.02& (xd)& (oe)\\
77& NGC 3516 &	0.0088& 43.26& 42.72& 21.55& \nodata&  41.54$\pm$ 0.01&  40.92$\pm$ 0.01&  $-$12.32$\pm$ 0.01&  (4.9)&  \nodata& (xd)& (oc)\\
78& RX J1127.2+1909 &	0.1055& 44.79& 43.84&  0.00& \nodata&  43.02$\pm$ 0.14&  43.02$\pm$ 0.11&  $-$12.43$\pm$ 0.11&  $-$13.03$\pm$ 0.01&  $-$13.42$\pm$ 0.03& (xa)& (oe)\\
79& NGC 3783 &	0.0097& 43.53& 43.30& 21.76& 0.278&  \nodata&  41.47$\pm$ 0.10&  $-$11.85$\pm$ 0.10&  \nodata&  \nodata& (xd)& (oa)\\
80& SBS 1136+594 &	0.0601& 44.33& 43.82&  0.00& \nodata&  42.70$\pm$ 0.06&  42.55$\pm$ 0.01&  $-$12.39$\pm$ 0.01&  $-$12.92$\pm$ 0.01&  $-$13.45$\pm$ 0.02& (xa)& (oe)\\
81& UGC 06728 &	0.0065& 42.54& 41.94& 20.00& \nodata&  40.22$\pm$ 0.01&  40.22$\pm$ 0.01&  $-$12.75$\pm$ 0.01&  $-$12.34$\pm$ 0.01&  $-$12.80$\pm$ 0.01& (xd)& (oe)\\
82& 2MASX J11454045--1827149 &	0.0330& 43.98& 43.64&  0.00& \nodata&  42.19$\pm$ 0.43&  42.19$\pm$ 0.10&  $-$12.21$\pm$ 0.10&  $-$12.91$\pm$ 0.10&  $-$13.36$\pm$ 0.10& (xa)& (oa)\\
83& CGCG 041--020 &	0.0360& 43.88& 43.39& 23.03& 0.009&  41.17$\pm$ 0.07&  40.38$\pm$ 0.01&  $-$14.10$\pm$ 0.01&  $-$13.96$\pm$ 0.01&  $-$14.70$\pm$ 0.02& (xd)& (oe)\\
85& NGC 4051 &	0.0023& 41.74& 41.33& 20.46& \nodata&  40.41$\pm$ 0.56&  40.20$\pm$ 0.18&  $-$11.87$\pm$ 0.18&  $-$11.97$\pm$ 0.01&  $-$12.52$\pm$ 0.18& (xc)& (oe)\\
86& Ark 347 &	0.0224& 43.42& 42.90& 23.36& 0.016&  41.53$\pm$ 0.22&  41.53$\pm$ 0.19&  $-$12.52$\pm$ 0.19&  $-$13.08$\pm$ 0.01&  $-$13.31$\pm$ 0.04& (xa)& (oe)\\
87& NGC 4102 &	0.0028& 41.62& 41.41& 24.30& \nodata&  40.72$\pm$ 0.02&  38.78$\pm$ 0.01&  $-$13.46$\pm$ 0.01&  $-$12.48$\pm$ 0.01&  $-$13.62$\pm$ 0.01& (xj)& (od)\\
88& NGC 4138 &	0.0030& 41.62& 41.23& 22.90& 0.012&  38.95$\pm$ 0.01&  38.95$\pm$ 0.01&  $-$13.35$\pm$ 0.01&  $-$13.33$\pm$ 0.01&  $-$13.72$\pm$ 0.01& (xd)& (od)\\
89& NGC 4151 &	0.0033& 42.96& 42.58& 22.73& 0.041&  41.87$\pm$ 0.24&  41.87$\pm$ 0.01&  $-$10.51$\pm$ 0.01&  $-$11.12$\pm$ 0.01&  $-$11.40$\pm$ 0.08& (xd)& (oe)\\
90& Mrk 766 &	0.0129& 42.94& 42.67& 21.72& \nodata&  42.03$\pm$ 0.13&  41.79$\pm$ 0.01&  $-$11.78$\pm$ 0.01&  $-$12.10$\pm$ 0.04&  $-$12.66$\pm$ 0.02& (xc)& (oe)\\
91& NGC 4388 &	0.0084& 43.60& 43.17& 23.53& 0.011&  41.29$\pm$ 0.13&  41.29$\pm$ 0.11&  $-$11.90$\pm$ 0.11&  $-$12.33$\pm$ 0.01&  $-$12.80$\pm$ 0.03& (xk)& (oe)\\
92& NGC 4395 &	0.0011& 40.81& 40.64& 22.52& 0.322&  39.02$\pm$ 0.01&  38.93$\pm$ 0.01&  $-$12.49$\pm$ 0.01&  $-$12.81$\pm$ 0.01&  $-$13.32$\pm$ 0.01& (xd)& (oe)\\
94& NGC 4507 &	0.0118& 43.78& 43.54& 23.54& 0.029&  42.12$\pm$ 0.43&  41.76$\pm$ 0.10&  $-$11.73$\pm$ 0.10&  $-$12.10$\pm$ 0.10&  $-$12.70$\pm$ 0.10& (xd)& (oa)\\
95$^{**}$& ESO 506-G027 &	0.0250& 44.28& 43.95& 23.92& 0.002& $>$41.14&  40.96$\pm$ 0.20&  $-$13.19$\pm$ 0.20&  $-$13.67$\pm$ 0.20& $<$$-$14.20& (xl)& (oa)\\
96& XSS J12389--1614 &	0.0366& 44.26& 43.35& 22.63& 0.043&  41.88$\pm$ 0.85&  41.87$\pm$ 0.20&  $-$12.63$\pm$ 0.20&  $-$12.81$\pm$ 0.20&  $-$13.29$\pm$ 0.20& (xa)& (oa)\\
97& NGC 4593 &	0.0090& 43.21& 42.81& 20.49& \nodata&  41.04$\pm$ 0.01&  40.50$\pm$ 0.01&  $-$12.76$\pm$ 0.01&  (4.6)&  \nodata& (xd)& (oc)\\
100& SBS 1301+540 &	0.0299& 43.72& 43.09& 20.60& \nodata&  41.25$\pm$ 0.21&  41.25$\pm$ 0.01&  $-$13.06$\pm$ 0.01&  $-$13.54$\pm$ 0.07&  $-$13.87$\pm$ 0.01& (xc)& (oe)\\
102$^{**}$& NGC 4992 &	0.0251& 43.83& 43.17& 23.75& $<$0.003&  39.85$\pm$ 0.16&  39.85$\pm$ 0.01&  $-$14.30$\pm$ 0.01&  $-$14.50$\pm$ 0.02&  $-$14.82$\pm$ 0.05& (xm)& (od)\\
103& MCG --03-34-064 &	0.0165& 43.46& 43.95& 23.61& 0.039&  42.55$\pm$ 0.43&  42.21$\pm$ 0.10&  $-$11.57$\pm$ 0.10&  $-$12.13$\pm$ 0.10&  $-$12.73$\pm$ 0.10& (xd)& (oa)\\
105& MCG --06-30-015 &	0.0077& 43.00& 42.74& 21.28& \nodata&  \nodata&  40.23$\pm$ 0.20&  $-$12.89$\pm$ 0.20&  \nodata&  \nodata& (xd)& (oa)\\
106& NGC 5252 &	0.0230& 43.90& 43.39& 22.64& 0.038&  40.79$\pm$ 0.01&  40.37$\pm$ 0.01&  $-$13.71$\pm$ 0.01&  $-$13.32$\pm$ 0.01&  $-$13.94$\pm$ 0.01& (xd)& (oe)\\
108& IC 4329A &	0.0160& 44.24& 43.84& 21.79& \nodata&  41.62$\pm$ 0.19&  41.02$\pm$ 0.04&  $-$12.74$\pm$ 0.04&  $-$13.09$\pm$ 0.04&  $-$13.77$\pm$ 0.04& (xd)& (od)\\
109& Mrk 279 &	0.0304& 43.97& 43.42& 20.11& \nodata&  41.76$\pm$ 0.21&  41.68$\pm$ 0.01&  $-$12.65$\pm$ 0.01&  $-$13.01$\pm$ 0.01&  $-$13.51$\pm$ 0.07& (xd)& (od)\\
110& NGC 5506 &	0.0062& 43.30& 42.91& 22.44& 0.011&  41.89$\pm$ 0.11&  41.03$\pm$ 0.08&  $-$11.90$\pm$ 0.08&  $-$11.99$\pm$ 0.01&  $-$12.76$\pm$ 0.03& (xd)& (oe)\\
112& NGC 5548 &	0.0172& 43.59& 43.14& 20.85& \nodata&  42.13$\pm$ 0.06&  42.13$\pm$ 0.02&  $-$11.69$\pm$ 0.02&  $-$12.37$\pm$ 0.01&  $-$12.74$\pm$ 0.02& (xd)& (oe)\\
113& ESO 511-G030 &	0.0224& 43.73& 43.41& 20.99& \nodata&  \nodata&  40.62$\pm$ 0.20&  $-$13.44$\pm$ 0.20&  \nodata&  \nodata& (xd)& (oa)\\
115& NGC 5728 &	0.0093& 43.23& 43.03& 24.14& 0.007&  41.96$\pm$ 0.43&  41.52$\pm$ 0.10&  $-$11.76$\pm$ 0.10&  $-$12.23$\pm$ 0.10&  $-$12.85$\pm$ 0.10& (xm)& (oa)\\
116& Mrk 841 &	0.0364& 44.20& 43.87& 21.34& \nodata&  41.64$\pm$ 0.04&  41.64$\pm$ 0.01&  $-$12.85$\pm$ 0.01&  $-$13.37$\pm$ 0.01&  $-$13.79$\pm$ 0.01& (xc)& (od)\\
117& Mrk 290 &	0.0296& 43.79& 42.93& 21.18& \nodata&  41.71$\pm$ 0.50&  41.59$\pm$ 0.01&  $-$12.71$\pm$ 0.01&  $-$13.20$\pm$ 0.01&  $-$13.72$\pm$ 0.17& (xd)& (od)\\
118& Mrk 1498 &	0.0547& 44.50& 44.05& 23.10& 0.016&  42.43$\pm$ 0.31&  42.43$\pm$ 0.20&  $-$12.42$\pm$ 0.20&  $-$13.18$\pm$ 0.01&  $-$13.05$\pm$ 0.08& (xf)& (oe)\\
120$^{**}$& NGC 6240 &	0.0245& 43.81& 44.16& 24.25& $<$0.005&  42.12$\pm$ 0.12&  40.71$\pm$ 0.02&  $-$13.43$\pm$ 0.02&  $-$12.73$\pm$ 0.01&  $-$13.69$\pm$ 0.04& (xa)& (oe)\\
124& 1RXS J174538.1+290823 &	0.1113& 45.09& 44.37&  0.00& \nodata&  \nodata&  42.75$\pm$ 0.03&  $-$12.75$\pm$ 0.03&  \nodata&  \nodata& (xa)& (oe)\\
125& 3C 382 &	0.0579& 44.81& 44.67& 20.11& \nodata&  42.31$\pm$ 0.08&  41.70$\pm$ 0.01&  $-$13.20$\pm$ 0.01&  $-$13.36$\pm$ 0.01&  $-$14.04$\pm$ 0.03& (xc)& (od)\\
126$^{**}$& ESO 103-035 &	0.0133& 43.58& 43.38& 23.33& 0.001&  42.20$\pm$ 0.85&  40.87$\pm$ 0.20&  $-$12.73$\pm$ 0.20&  $-$12.80$\pm$ 0.20&  $-$13.73$\pm$ 0.20& (xd)& (oa)\\
127& 3C 390.3 &	0.0561& 44.88& 44.52& 21.08& \nodata&  42.96$\pm$ 0.38&  42.72$\pm$ 0.01&  $-$12.15$\pm$ 0.01&  $-$12.51$\pm$ 0.13&  $-$13.06$\pm$ 0.02& (xd)& (od)\\
129& NGC 6814 &	0.0052& 42.57& 42.22& 20.76& \nodata&  \nodata&  40.17$\pm$ 0.10&  $-$12.61$\pm$ 0.10&  \nodata&  \nodata& (xc)& (oa)\\
133& NGC 6860 &	0.0149& 43.39& 42.89& 21.00& \nodata&  \nodata&  40.93$\pm$ 0.10&  $-$12.76$\pm$ 0.10&  \nodata&  \nodata& (xn)& (oa)\\
136& 4C +74.26 &	0.1040& 45.14& 44.87& 21.25& \nodata&  43.37$\pm$ 0.61&  43.13$\pm$ 0.17&  $-$12.31$\pm$ 0.17&  $-$12.83$\pm$ 0.01&  $-$13.39$\pm$ 0.20& (xo)& (oe)\\
137& Mrk 509 &	0.0344& 44.43& 44.08& 20.18& \nodata&  \nodata&  42.17$\pm$ 0.20&  $-$12.26$\pm$ 0.20&  \nodata&  \nodata& (xd)& (oa)\\
138& IC 5063 &	0.0114& 43.31& 43.08& 23.40& 0.009&  42.26$\pm$ 0.43&  41.58$\pm$ 0.10&  $-$11.88$\pm$ 0.10&  $-$12.20$\pm$ 0.10&  $-$12.91$\pm$ 0.10& (xh)& (oa)\\
139& 2MASX J21140128+8204483 &	0.0840& 44.80& 44.35&  0.00& \nodata& $>$43.60&  42.89$\pm$ 0.01&  $-$12.35$\pm$ 0.01&  $-$12.54$\pm$ 0.04& $<$$-$13.25& (xa)& (od)\\
144& UGC 11871 &	0.0266& 43.80& 43.26& 22.32& 0.016&  42.56$\pm$ 0.01&  41.45$\pm$ 0.01&  $-$12.76$\pm$ 0.01&  $-$12.25$\pm$ 0.01&  $-$13.10$\pm$ 0.01& (xa)& (oe)\\
145$^{**}$& NGC 7172 &	0.0087& 43.32& 42.90& 22.91& 0.001&  39.94$\pm$ 0.85&  39.94$\pm$ 0.20&  $-$13.28$\pm$ 0.20&  $-$13.55$\pm$ 0.20&  $-$13.75$\pm$ 0.20& (xd)& (oa)\\
146& NGC 7213 &	0.0058& 42.59& 41.85& 20.40& \nodata&  40.23$\pm$ 0.85&  40.23$\pm$ 0.20&  $-$12.64$\pm$ 0.20&  $-$12.39$\pm$ 0.20&  $-$12.83$\pm$ 0.20& (xd)& (oa)\\
147& NGC 7314 &	0.0048& 42.45& 41.96& 21.60& \nodata&  40.25$\pm$ 0.43&  39.78$\pm$ 0.10&  $-$12.93$\pm$ 0.10&  $-$13.21$\pm$ 0.10&  $-$13.85$\pm$ 0.10& (xd)& (oa)\\
148$^{**}$& NGC 7319 &	0.0225& 43.68& 43.40& 23.82& 0.004&  40.72$\pm$ 0.20&  40.72$\pm$ 0.03&  $-$13.34$\pm$ 0.03&  $-$13.68$\pm$ 0.01&  $-$13.86$\pm$ 0.07& (xa)& (oe)\\
149& 3C 452 &	0.0811& 44.73& 43.83& 23.36& 0.064&  42.09$\pm$ 0.09&  40.98$\pm$ 0.01&  $-$14.23$\pm$ 0.01&  $-$14.22$\pm$ 0.01&  $-$15.07$\pm$ 0.03& (xd)& (oe)\\
151& MR 2251--178 &	0.0640& 45.03& 44.60& 21.45& \nodata&  42.87$\pm$ 0.43&  42.33$\pm$ 0.10&  $-$12.66$\pm$ 0.10&  $-$12.89$\pm$ 0.10&  $-$13.55$\pm$ 0.10& (xd)& (oa)\\
152& NGC 7469 &	0.0163& 43.70& 42.97& 20.61& \nodata&  43.14$\pm$ 0.01&  41.70$\pm$ 0.01&  $-$12.08$\pm$ 0.01&  $-$11.89$\pm$ 0.01&  $-$12.85$\pm$ 0.01& (xc)& (ob)\\
153& Mrk 926 &	0.0469& 44.45& 44.19& 20.54& \nodata&  42.66$\pm$ 0.03&  42.66$\pm$ 0.01&  $-$12.05$\pm$ 0.01&  $-$12.68$\pm$ 0.01&  $-$13.05$\pm$ 0.01& (xd)& (oe)\\
154& NGC 7582 &	0.0052& 42.61& 42.65& 23.80& 0.033&  41.26$\pm$ 0.43&  40.38$\pm$ 0.10&  $-$12.39$\pm$ 0.10&  $-$12.10$\pm$ 0.10&  $-$12.88$\pm$ 0.10& (xp)& (oa)
\enddata
\tablecomments{
This table summarizes X-ray and optical emission line (\oiii, \ha, \hb) properties
of 95 {\it Swift}/BAT 9-month AGNs in \citet{tue08} excluding Cen A, blazars,
and those at low Galactic latitudes ($|b|<15^\circ$). Columns::
(1) source no.\ in \cite{tue08}; (2) object name; (3) redshift;
(4) 9-month averaged 14--195 keV luminosity calculated from the observed flux;
(5) absorption-corrected 2--10 keV luminosity of the transmitted component averaged for 70 months;
(6) X-ray absorption hydrogen column density ($0.00$ means $N_{\rm H}=0$);
(7) soft X-ray scattering fraction;
(8) \oiii\ luminosity corrected for extinction based on the Balmer
 decrement;
(9) observed \oiii\ luminosity (with no extinction correction);
(10) \oiii\ flux;
(11) narrow \ha\ flux (number in parenthesis refers to \ha / \hb\ flux ratio);
(12) narrow \hb\ flux;
(13) reference for the X-ray spectra (Columns 6 and 7):
(xa) \citet{ric15}, (xb) \citet{nog10}, (xc) \citet{tue08}, (xd)
 \citet{win09a}, (xe) \citet{egu11}, (xf) \citet{egu09}, (xg) \citet{ris09}
(xh) \citet{tur09}, (xi) \citet{awa08}, (xj) \citet{gon11}, (xk)
 \citet{shi08}, (xl) \citet{win09b}, (xm) \citet{com10}, (xn)
 \citet{win10b}, (xo) \citet{bal05}, (xp) bia \citet{bia09}.
(14) reference for the optical line fluxes (Columns 8--12):
(oa) this work, (ob) \citet{dah88}, (oc) \citet{mul94}, (od) \citet{kos15}, (oe) \citet{win10a}, (of) \citet{bas99}, (og) \citet{xu99}, (oh) \citet{lan07}.
Columns (1)--(4) are taken from \citet{tue08} except for the revised redshift of 
no.~53 (2MASX J06403799--4321211). Column (5) is taken from C. Ricci et al.\ (2015, in preparation). All luminosities are calculated from the redshift
given in column (3) with ($H_0$, $\Omega_{\rm m}$, $\Omega_{\lambda}$)
= (70 km s$^{-1}$ Mpc$^{-1}$, 0.3, 0.7).
}
\end{deluxetable}
\clearpage
%\end{landscape}

\tablefontsize{\footnotesize}
\begin{deluxetable}{cccccccccc}
\tabletypesize{\footnotesize}
\tablecaption{
Correlation Properties between Different Luminosities}
\label{tbl-2}
\tablewidth{0pt}
\tablenum{2}
\tablehead{
\colhead{Y} & \colhead{X} & \colhead{Sample} & \colhead{$N$} &
\colhead{$\rho_{L}$} & \colhead{$\rho_{f}$} & \colhead{$P_L$} & \colhead{$P_f$} & \colhead{$a$} & \colhead{$b$}\\
\colhead{(1)} & \colhead{(2)} & \colhead{(3)} & \colhead{(4)} &
\colhead{(5)} & \colhead{(6)} & \colhead{(7)} & \colhead{(8)} & \colhead{(9)} & \colhead{(10)}\\
}
\startdata
    &       & All &  101& 0.672 &0.419 & $1.4\times10^{-14}$ & $1.3\times10^{-5}$ & $-11.0\pm 2.9$ & $1.20\pm0.07$ \\
\Lo & \Lxh & Type 1 &  48& 0.781 &0.240 & $5.9\times10^{-11}$ & $1.0\times10^{-1}$ & $-8.0\pm 3.8$ & $1.13\pm0.09$ \\
    &       & Type 2 &  53& 0.461 &0.548 & $5.1\times10^{-4}$ & $2.2\times10^{-5}$ & $-10.3\pm 3.7$ & $1.18\pm0.09$ \\
\cline{1-10}
    &       & All &  100& 0.630 &0.375 & $2.1\times10^{-12}$ & $1.2\times10^{-4}$ & $-10.0\pm 2.9$ & $1.18\pm0.07$ \\
\Lo & \Lxs & Type 1 &  48& 0.761 &0.239 & $3.4\times10^{-10}$ & $1.0\times10^{-1}$ & $-5.5\pm 3.6$ & $1.08\pm0.09$ \\
    &       & Type 2 &  52& 0.447 &0.488 & $9.1\times10^{-4}$ & $2.4\times10^{-4}$ & $-10.2\pm 4.0$ & $1.19\pm0.10$ \\
\cline{1-10}
    &       & All &  76& 0.594 &0.475 & $1.5\times10^{-8}$ & $1.5\times10^{-5}$ & $-9.8\pm 4.0$ & $1.18\pm0.09$ \\
\Locor & \Lxh & Type 1 &  31& 0.705 &0.341 & $9.5\times10^{-6}$ & $6.0\times10^{-2}$ & $-5.4\pm 4.2$ & $1.08\pm0.10$ \\
    &       & Type 2 &  45& 0.421 &0.571 & $4.0\times10^{-3}$ & $4.2\times10^{-5}$ & $-12.0\pm 5.5$ & $1.23\pm0.13$ \\
\cline{1-10}
    &       & All &  75& 0.619 &0.516 & $3.2\times10^{-9}$ & $2.2\times10^{-6}$ & $-8.5\pm 3.7$ & $1.16\pm0.09$ \\
\Locor & \Lxs & Type 1 &  31& 0.716 &0.463 & $5.9\times10^{-6}$ & $8.8\times10^{-3}$ & $-1.6\pm 3.5$ & $1.01\pm0.09$ \\
    &       & Type 2 &  44& 0.503 &0.552 & $5.0\times10^{-4}$ & $1.0\times10^{-4}$ & $-12.8\pm 5.3$ & $1.26\pm0.13$ \\
\cline{1-10}
    &       & All &  82& 0.608 &0.352 & $1.4\times10^{-9}$ & $1.2\times10^{-3}$ & $-3.0\pm 3.1$ & $1.02\pm0.08$ \\
\Lha & \Lxs & Type 1 &  32& 0.739 &0.128 & $1.3\times10^{-6}$ & $4.8\times10^{-1}$ & $ 3.9\pm 3.9$ & $0.87\pm0.09$ \\
    &       & Type 2 &  50& 0.542 &0.514 & $4.9\times10^{-5}$ & $1.3\times10^{-4}$ & $-4.0\pm 4.2$ & $1.04\pm0.10$ \\
\cline{1-10}
    &       & All &  82& 0.934 &0.851 & $2.1\times10^{-37}$ & $4.6\times10^{-24}$ & $-7.6\pm 2.1$ & $1.19\pm0.05$ \\
\Lo & \Lha & Type 1 &  32& 0.937 &0.799 & $3.1\times10^{-15}$ & $4.1\times10^{-8}$ & $-6.2\pm 2.7$ & $1.16\pm0.07$ \\
    &       & Type 2 &  50& 0.886 &0.877 & $1.2\times10^{-17}$ & $6.4\times10^{-17}$ & $-7.6\pm 3.1$ & $1.03\pm0.11$ 
\enddata
\tablecomments{
Columns: (1) $Y$ variable; (2) $X$ variable; (3) AGN type; (4) number of
 sample; (5) Spearman's rank coefficient for luminosity--luminosity
 correlation ($\rho_{\rm L}$); (6) Spearman's rank coefficient for
 flux--flux correlation ($\rho_{\rm f}$); (7) Student's $t$-null
 significance level for luminosity--luminosity correlation ($P_{\rm L}$); (8)
 Student's $t$-null significance level for flux--flux correlation
 ($P_{\rm f}$); (9) regression intercept ($a$) and its 1$\sigma$
 uncertainty; (10) slope ($b$) and its 1$\sigma$ uncertainty. Equation is represented as $Y=a+bX$.
}
\end{deluxetable}

\tablefontsize{\footnotesize}
\begin{deluxetable}{cccccc}
\tabletypesize{\footnotesize}
\tablecaption{Summary of Luminosity Ratios}
\label{tbl-3}
\tablewidth{0pt}
\tablenum{3}
\tablehead{
\colhead{Luminosity} &\colhead{Sample} & \colhead{$N$} & \colhead{$<r>$} &
 \colhead{$\sigma$} & \colhead{$<L_{\rm X}>$} \\
\colhead{Ratio} & \colhead{(1)} & \colhead{(2)} & \colhead{(3)} &
 \colhead{(4)} & \colhead{(5)}
}
\startdata
       &  All & 102 & $-2.45\pm0.06$ & $0.60\pm0.05$ & 43.71 \\
       &  Type 1 & 48 & $-2.27\pm0.07$ & $0.47\pm0.05$ & 43.85 \\
\LoLxh &  Type 2 & 54 & $-2.62\pm0.10$ & $0.66\pm0.07$ & 43.60 \\
       &  Type 2 ($f_{\rm scat}<0.5\%$) & 12 & $-3.13\pm0.10$ & $0.33\pm0.08$ & 43.92 \\
       &  Type 2 ($i_{\rm host}>80^\circ$) & 10 & $-2.43\pm0.13$ & $0.41\pm0.10$ & 43.24 \\
\cline{1-6}
       &  All & 101 & $-1.99\pm0.07$ & $0.63\pm0.05$ & 43.24 \\
       &  Type 1 & 48 & $-1.78\pm0.08$ & $0.51\pm0.06$ & 43.36 \\
\LoLxs &  Type 2 & 53 & $-2.19\pm0.10$ & $0.67\pm0.07$ & 43.14 \\
       &  Type 2 ($f_{\rm scat}<0.5\%$) & 12 & $-2.85\pm0.09$ & $0.31\pm0.07$ & 43.62 \\
       &  Type 2 ($i_{\rm host}>80^\circ$) & 10 & $-2.01\pm0.17$ & $0.53\pm0.13$ & 42.81 \\
\cline{1-6}
       &  All & 77 & $-1.94\pm0.08$ & $0.69\pm0.06$ & 43.65 \\
       &  Type 1 & 31 & $-1.89\pm0.09$ & $0.48\pm0.07$ & 43.79 \\
\LocorLxh &  Type 2 & 46 & $-1.98\pm0.12$ & $0.80\pm0.09$ & 43.56 \\
       &  Type 2 ($f_{\rm scat}<0.5\%$) & 9 & $-2.44\pm0.28$ & $0.82\pm0.21$ & 43.87 \\
       &  Type 2 ($i_{\rm host}>80^\circ$) & 7 & $-1.63\pm0.23$ & $0.59\pm0.17$ & 43.06 \\
\cline{1-6}
       &  All & 76 & $-1.48\pm0.08$ & $0.69\pm0.06$ & 43.18 \\
       &  Type 1 & 31 & $-1.38\pm0.10$ & $0.51\pm0.07$ & 43.28 \\
\LocorLxs &  Type 2 & 45 & $-1.55\pm0.12$ & $0.78\pm0.09$ & 43.11 \\
       &  Type 2 ($f_{\rm scat}<0.5\%$) & 9 & $-2.16\pm0.24$ & $0.71\pm0.18$ & 43.58 \\
       &  Type 2 ($i_{\rm host}>80^\circ$) & 7 & $-1.20\pm0.29$ & $0.76\pm0.22$ & 42.62 \\
\cline{1-6}
       &  All & 83 & $-2.18\pm0.07$ & $0.61\pm0.05$ & 43.22 \\
       &  Type 1 & 32 & $-1.95\pm0.10$ & $0.54\pm0.07$ & 43.36 \\
\LhaLxs &  Type 2 & 51 & $-2.35\pm0.09$ & $0.60\pm0.07$ & 43.14 \\
       &  Type 2 ($f_{\rm scat}<0.5\%$) & 13 & $-2.97\pm0.08$ & $0.28\pm0.06$ & 43.49 \\
       &  Type 2 ($i_{\rm host}>80^\circ$) & 11 & $-2.28\pm0.18$ & $0.57\pm0.13$ & 42.73 \\
\cline{1-6}
       &  All & 83 & $ 0.23\pm0.04$ & $0.32\pm0.03$ & 41.06 \\
       &  Type 1 & 32 & $ 0.32\pm0.05$ & $0.27\pm0.04$ & 41.40 \\
\LoLha &  Type 2 & 51 & $ 0.16\pm0.05$ & $0.34\pm0.04$ & 40.85 \\
       &  Type 2 ($f_{\rm scat}<0.5\%$) & 12 & $ 0.15\pm0.09$ & $0.30\pm0.07$ & 40.63 \\
       &  Type 2 ($i_{\rm host}>80^\circ$) & 10 & $ 0.26\pm0.06$ & $0.17\pm0.05$ & 40.52
\enddata
\tablecomments{
Columns: (1) luminosity ratio; (2) AGN type; (3) number of objects; (4) average; (5) standard deviation; (6) mean luminosity value of the denominator in the sample.
}
\end{deluxetable}

\begin{figure}
\epsscale{0.45}
\includegraphics[angle=0,scale=0.8]{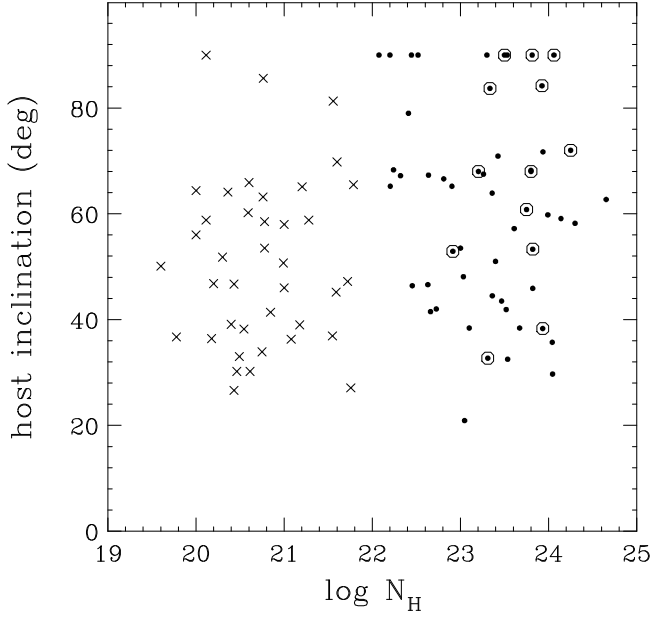}
\caption{Plot of host-galaxy inclination (\ihost) versus X-ray absorption
column density (\nh) for Sample~A (see Section~\ref{sec:catalog}).
The diagonal crosses, filled circles, filled+open circles correspond to
the \typeone\ (unabsorbed) AGNs, \typetwo\ (absorbed) AGNs, and
\typetwo\ AGNs with low scattering fractions. 
\label{fig1}}
\end{figure}

\begin{figure}
\includegraphics[angle=0,scale=0.8]{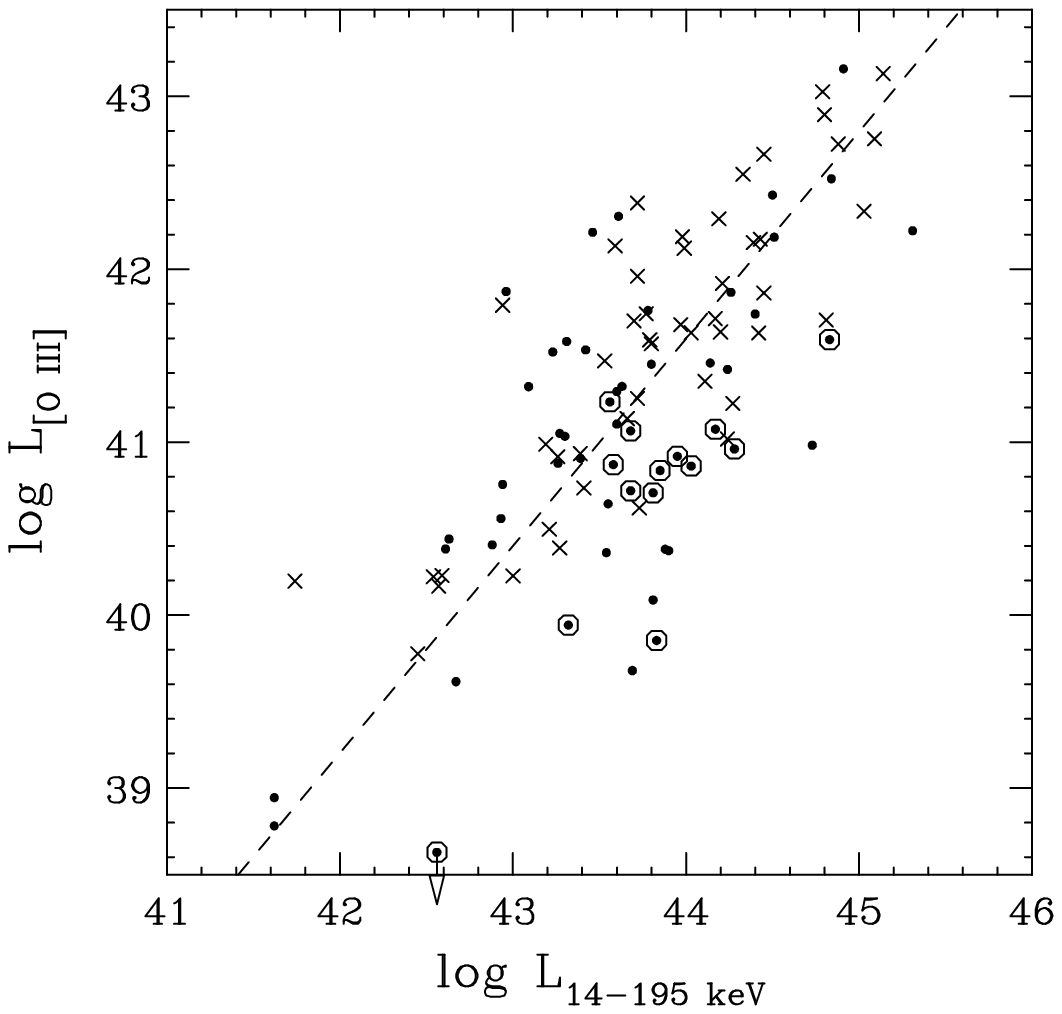}
\includegraphics[angle=0,scale=0.8]{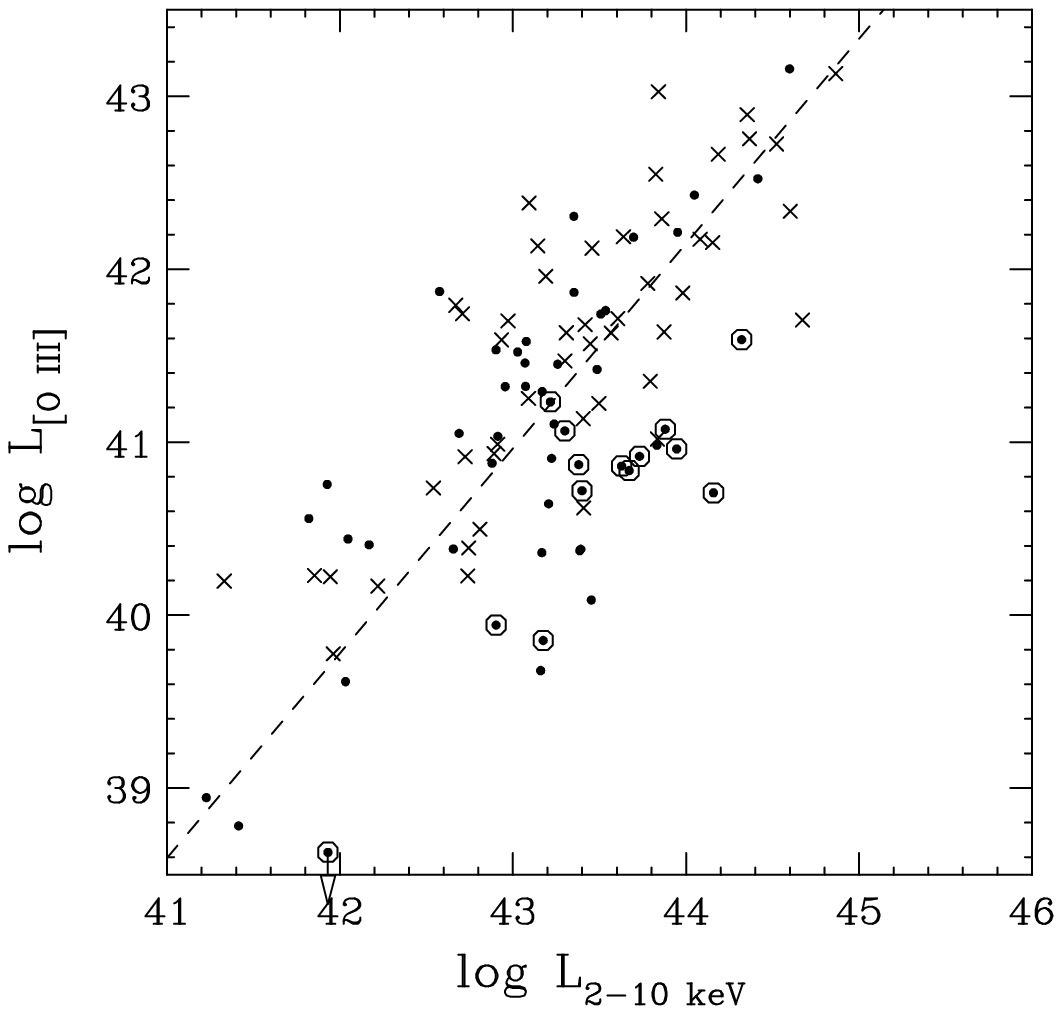}
\caption{{\it Left}: (a) Correlation between the observed X-ray luminosity in
the 14--195 keV band and observed \oiii\ luminosity for Sample~A.
{\it Right}: (b) Correlation between the intrinsic X-ray luminosity in
the 2--10 keV band and observed \oiii\ luminosity for Sample~A.
The lines are the best-fit regression lines obtained from all (\typeone\
 and \typetwo ) AGNs.
The symbols are the same as in Figure~1.
\label{fig2}}
\end{figure}

\begin{figure}
\includegraphics[angle=0,scale=0.8]{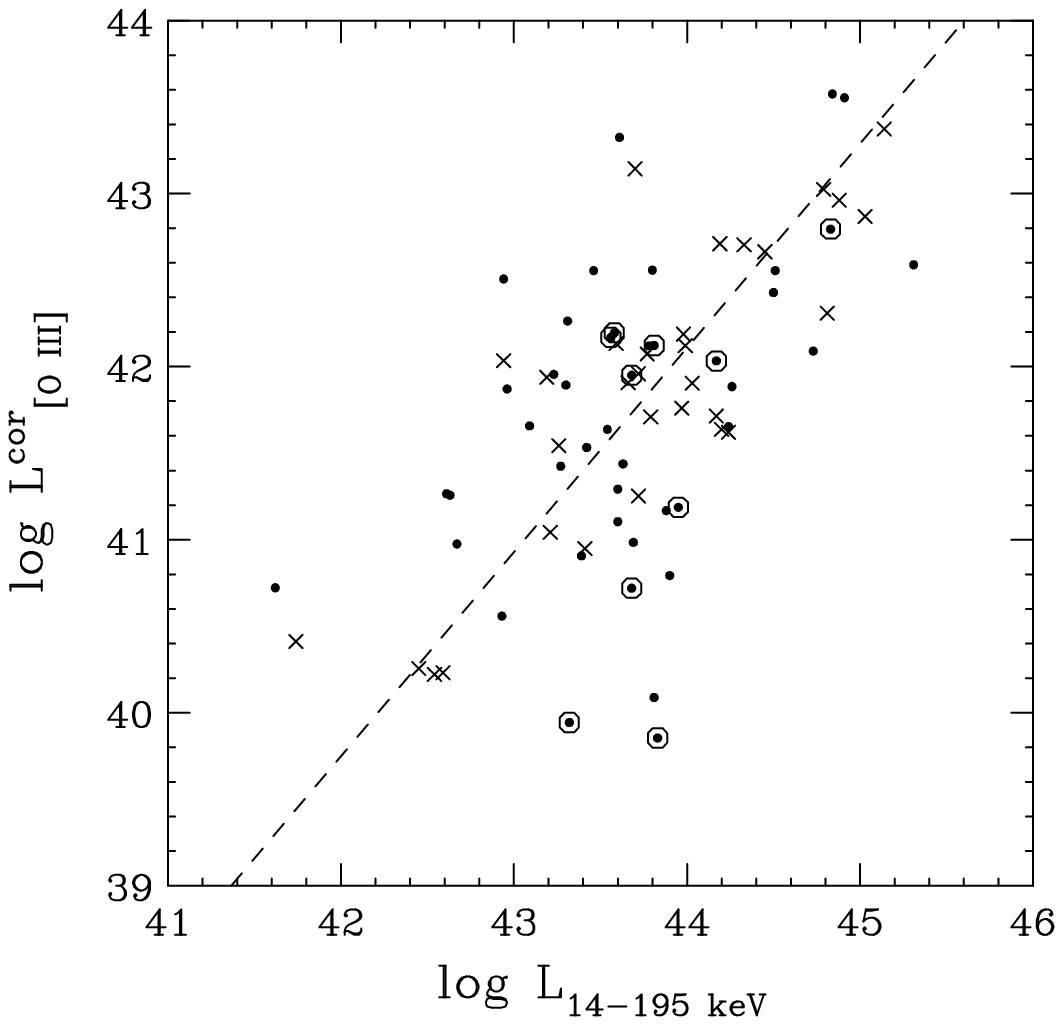}
\includegraphics[angle=0,scale=0.8]{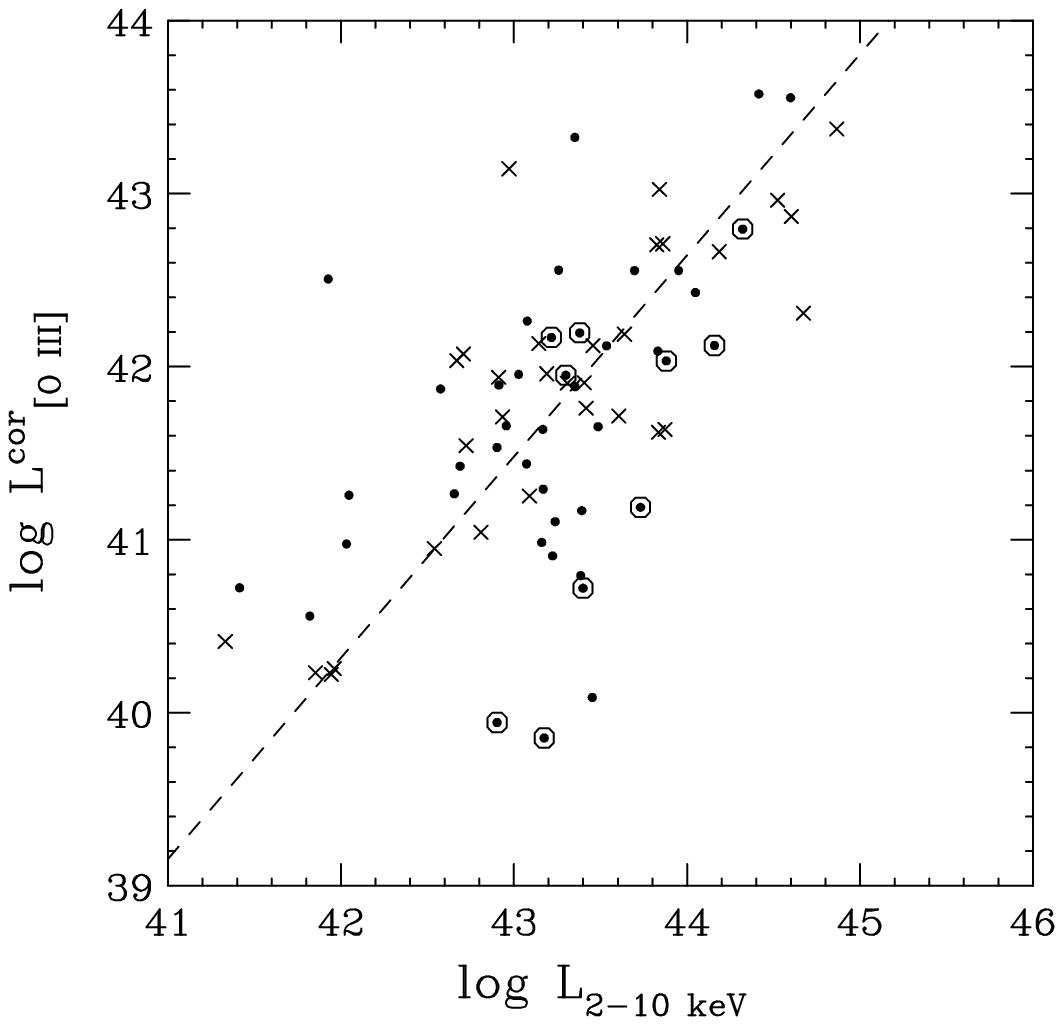}
\caption{
{\it Left}: (a) Correlation between the observed X-ray luminosity in
 the 14--195 keV band and extinction-corrected \oiii\ luminosity for Sample~B.
{\it Right}: (b) Correlation between the intrinsic X-ray luminosity in
 the 2--10 keV band and extinction-corrected \oiii\ luminosity for Sample~B.
The lines are the best-fit regression lines obtained from all (\typeone\
 and \typetwo ) AGNs.
The symbols are the same as in Figure~1.
\label{fig3}
}
\end{figure}

\begin{figure}
\includegraphics[angle=0,scale=0.8]{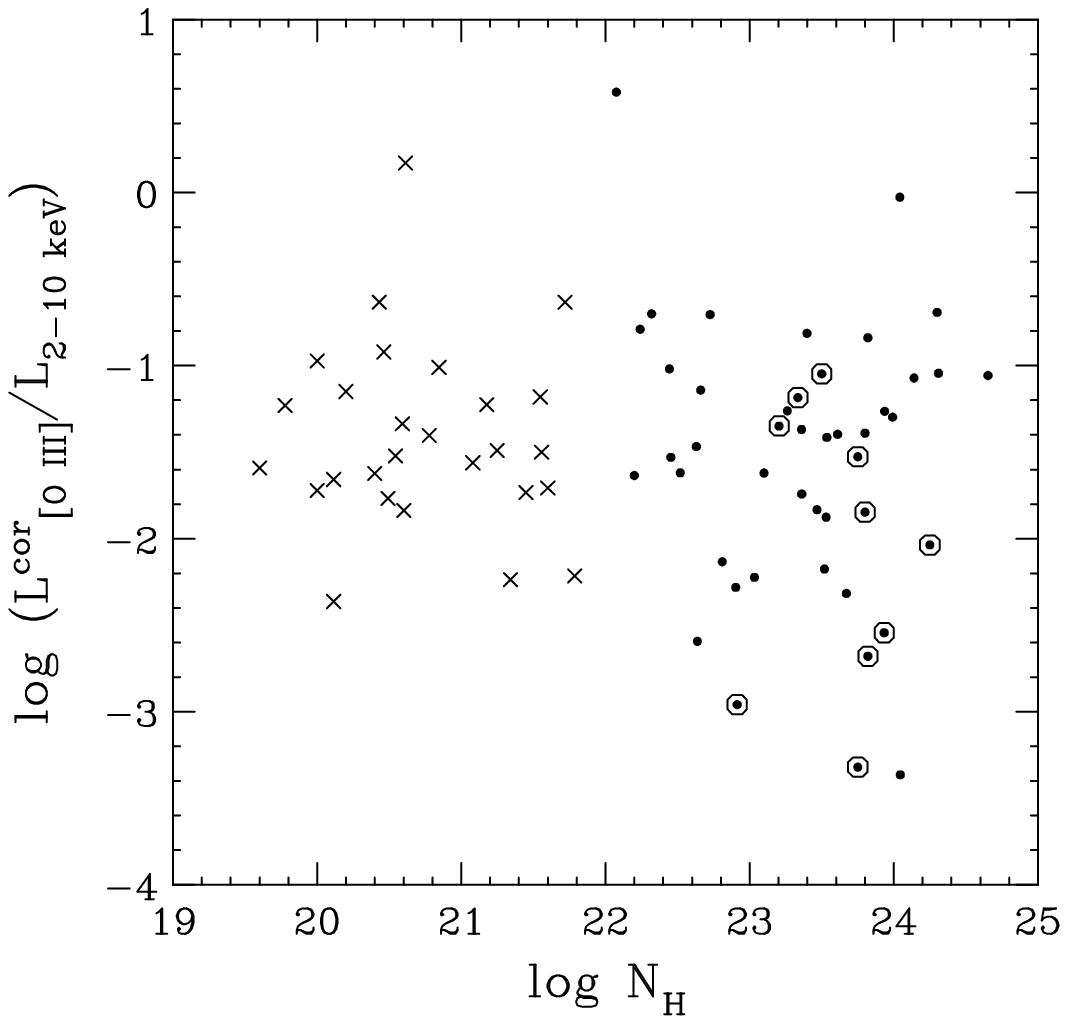}
\includegraphics[angle=0,scale=0.8]{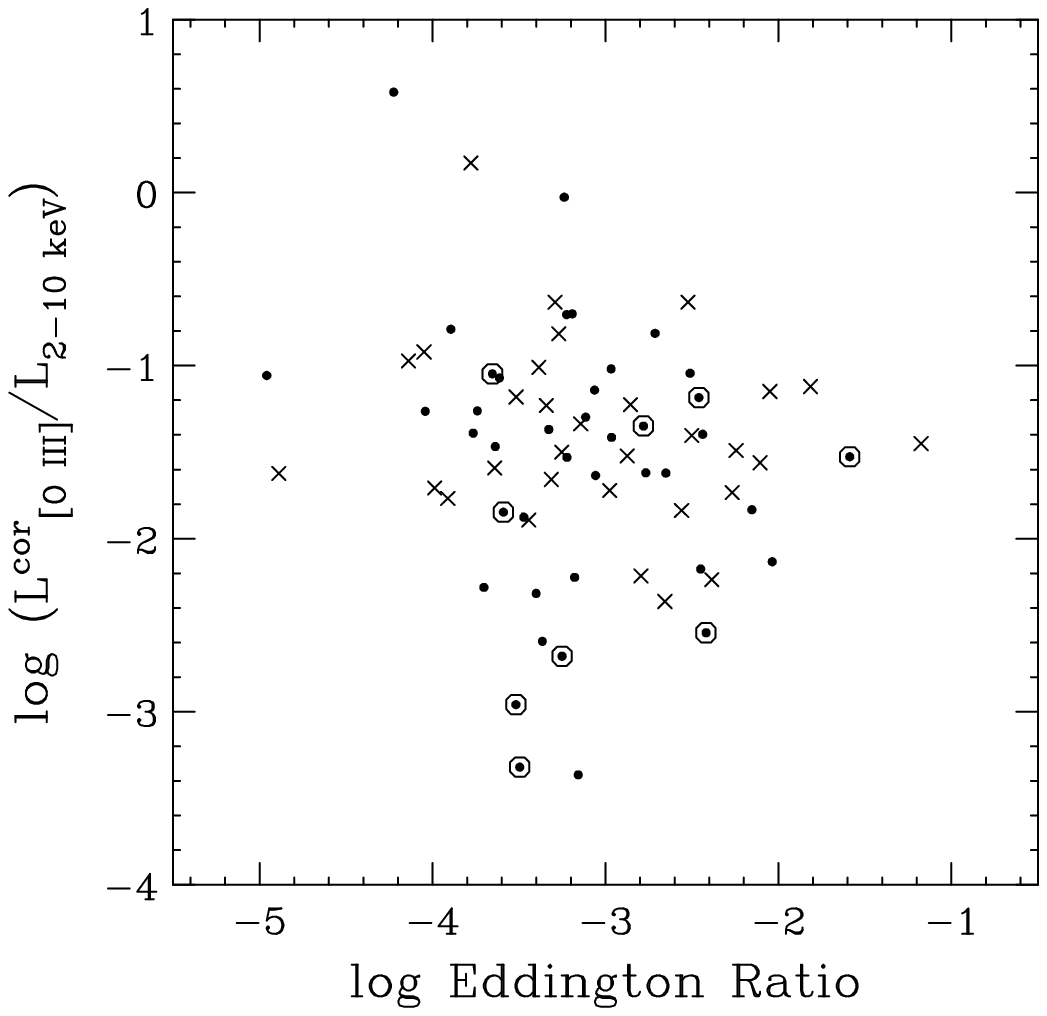}
\caption{
{\it Left}: (a) The ratio of the extinction-corrected \oiii\
 luminosity to the intrinsic 2--10 keV luminosity plotted against \nh\
 for Sample~B.
{\it Right}: (b) The ratio of the extinction-corrected \oiii\
 luminosity to the intrinsic 2--10 keV luminosity plotted against 
X-ray Eddington ratio (the 2--10 keV luminosity normalized by the Eddington
luminosity) for Sample~B.
The symbols are the same as in Figure~1.
\label{fig4}}
\end{figure}

\begin{figure}
\includegraphics[angle=0,scale=0.8]{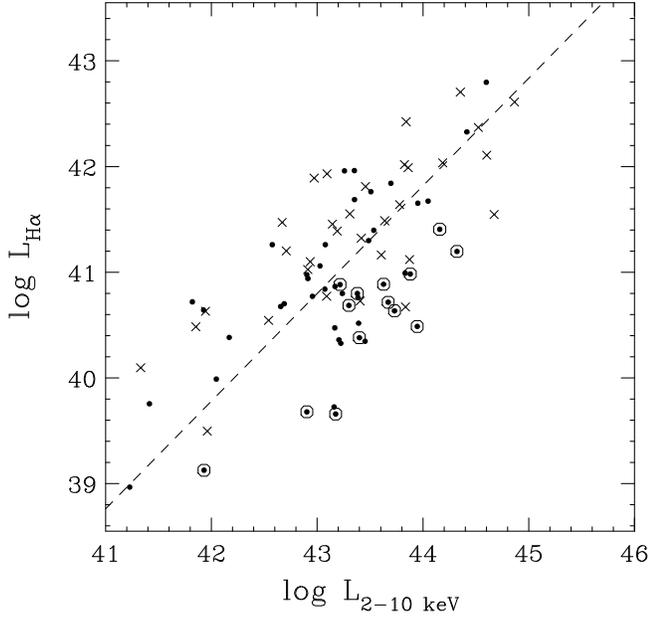}
\caption{
Correlation between the intrinsic X-ray luminosity in
the 2--10 keV band and H$\alpha$ luminosity for AGNs in Sample~A with available \ha\ fluxes.
The lines are the best-fit regression lines obtained from all (\typeone\
 and \typetwo ) AGNs.
The symbols are the same as in Figure~1.
\label{fig5}}
\end{figure}

\begin{figure}
\includegraphics[angle=0,scale=0.8]{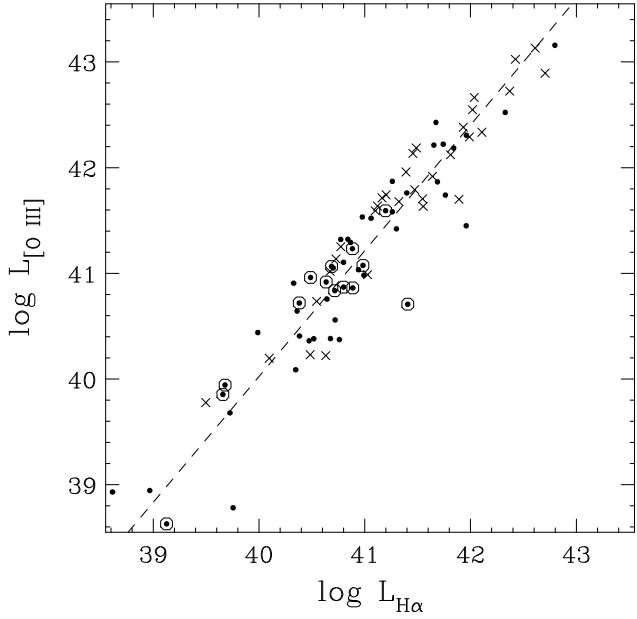}
\caption{
Correlation between the H$\alpha$ luminosity and 
observed \oiii\ luminosity for AGNs in Sample~A with available \ha\ fluxes.
The lines are the best-fit regression lines
obtained from all (\typeone\ and \typetwo ) AGNs. The symbols are the same as in Figure~1.
\label{fig6}}
\end{figure}

\begin{figure}
\includegraphics[angle=0,scale=0.5]{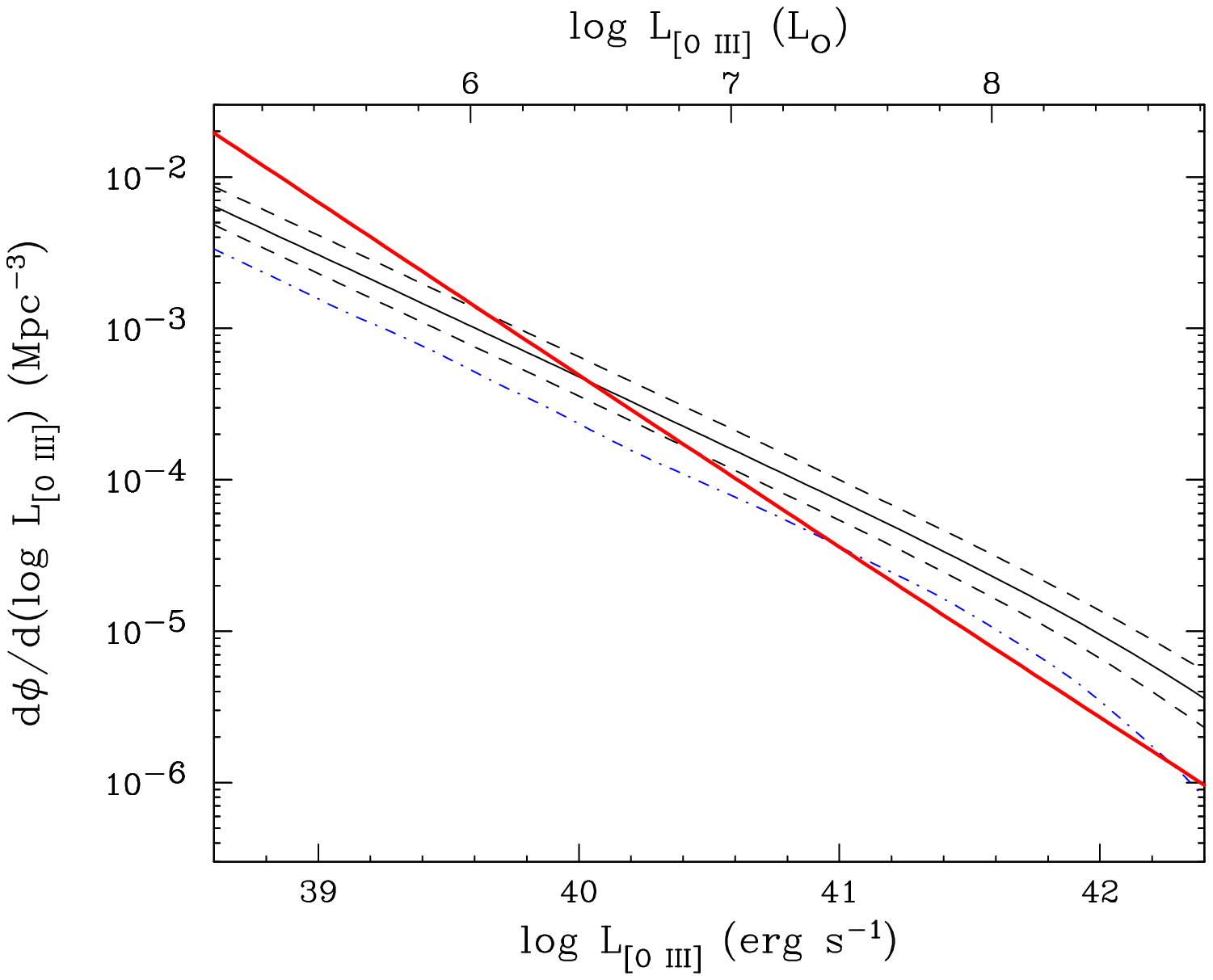}
\includegraphics[angle=0,scale=0.5]{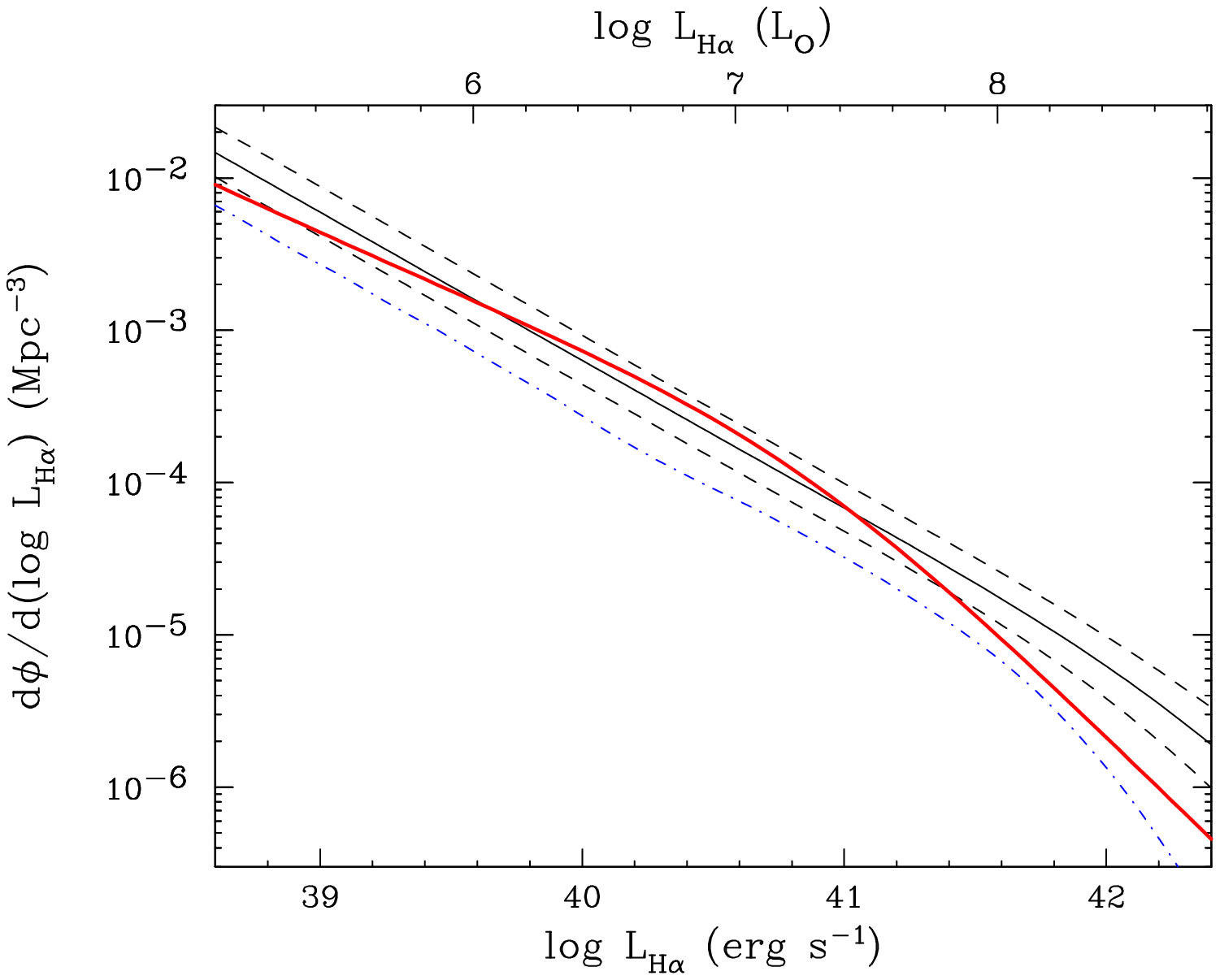}
\caption{
{\it Left}: (a) Comparison of \oiii\ and X-ray luminosity functions
 (LFs) of local AGNs. 
The thick solid curve (red) represents the observed \oiii\ LF from the SDSS. 
The solid curve (black) is a predicted \oiii\ LF from the X-ray LF 
in the \citet{ued14} model. The region surrounded by the two dashed curves (black) 
reflects the $1\sigma$ uncertainties in the mean and standard deviation in the 
\LoLxs\ ratio. The dot-dashed curve (blue) corresponds to the case with the standard
deviation is set to zero. The upper axis gives the \oiii\ luminosities in solar units.
{\it Right}: (b) Same as (a) but for \ha\ luminosity function.
\label{fig7}}
\end{figure}

\end{document}